\documentclass[10pt]{iopart}
\usepackage[ngerman,english]{babel}

\usepackage{graphicx}
\usepackage{subcaption}

\expandafter\let\csname equation*\endcsname\relax
\expandafter\let\csname endequation*\endcsname\relax
\usepackage{amsmath,amssymb}
\usepackage{bm}

\usepackage{letltxmacro}
\makeatletter
\let\oldr@@t\r@@t
\def\r@@t#1#2{%
	\setbox0=\hbox{$\oldr@@t#1{#2\,}$}\dimen0=\ht0
	\advance\dimen0-0.2\ht0
	\setbox2=\hbox{\vrule height\ht0 depth -\dimen0}%
	{\box0\lower0.4pt\box2}}
\LetLtxMacro{\oldsqrt}{\sqrt}
\renewcommand*{\sqrt}[2][\ ]{\oldsqrt[#1]{#2}}
\makeatother

\usepackage{mathtools}

\DeclarePairedDelimiter\floor{\lfloor}{\rfloor}
\newcommand{\IPP}{Max-Planck-Institut f\"ur Plasmaphysik, Boltzmannstr.~2, 85748 Garching, Germany}

\newcount\colveccount
\newcommand*\colvec[1]{
        \global\colveccount#1
        \begin{pmatrix}
        \colvecnext
}
\def\colvecnext#1{
        #1
        \global\advance\colveccount-1
        \ifnum\colveccount>0
                \\
                \expandafter\colvecnext
        \else
                \end{pmatrix}
        \fi
}

\newcommand{\grad}[2][]{\nabla_{#1} #2}

\newcommand{\dif}{\mathop{}\!\mathrm{d}}

\newcommand{\df}{\textrm{d}}

\newcommand{\mun}{\mu_\textnormal{0}}
\newcommand{\unitvecb}{\boldsymbol{b}}
\newcommand{\vpar}{v_{\parallel}}
\newcommand{\vpardot}{\dot{v}_{\parallel}}
\newcommand{\Apar}{A_{\parallel}}
\newcommand{\Aparh}{A_{\parallel}^{(h)}}
\newcommand{\Apars}{A_{\parallel}^{(s)}}
\newcommand{\bparstar}{B_{\parallel}^{*}}
\renewcommand{\vec}[1]{\boldsymbol{#1}}
\newcommand{\pder}[2][]{\frac{\partial#1}{\partial#2}}

\newcommand{\der}[2][]{\frac{\dif#1}{\dif#2}}

\newcommand{\Alfv}{Alfv\'en}

\newcommand{\omegaA}{\omega_\textnormal{A}}
\newcommand{\omegaci}{\omega_\textnormal{ci}}
\newcommand{\wci}{\omegaci}
\newcommand{\wa}{\omegaA}
\newcommand{\trm}[1]{\textrm{#1}}

\graphicspath{{images/}}

\begin{document}
\title{Global linear and nonlinear gyrokinetic modelling of Alfv\'en eigenmodes in ITER}
\author{T.~Hayward-Schneider$^1$, Ph.~Lauber$^1$, A.~Bottino$^1$, and Z.X.~Lu$^1$}
\address{$^1$ \IPP}
\ead{thomas.hayward@ipp.mpg.de}
\begin{abstract}
Linear and nonlinear modelling of \Alfv{}ic instabilities, most notably toroidal \Alfv{} eigenmodes (TAEs), obtained by using the global nonlinear electromagnetic gyrokinetic model of the code ORB5 are presented for the 15$\ $MA scenario of the ITER tokamak. Linear simulations show that elliptic \Alfv{} eigenmodes and odd-parity TAEs are only weakly damped but not excited by alpha particles, whose drive favours even-parity TAEs. Low mode number TAEs are found to be global, requiring global treatment.
Nonlinearly, even with double the nominal EP density, single mode simulations lead to saturation with negligible EP transport however multi-mode simulations predict that with double the nominal EP density, enhanced saturation and significant EP redistribution will occur.
\end{abstract}

	\maketitle
	\ioptwocol 

        \section{Introduction}
\label{sec:introduction}
Energetic particles, such as those born in fusion reactions, must be well confined if we can hope for self-heating `burning' plasmas in future tokamak reactors.
Already in present day machines, plasma heating from neutral beam injection (NBI) gives birth to typically sub-\Alfv{}ic ($v_\trm{injection} < v_\trm{A}$, where $v_\trm{A} = B/(\mu_0 n_\trm{i} m_\trm{i})$), which are able to interact with shear \Alfv{} waves (SAWs) and resonantly drive \Alfv{} eigenmodes (AEs).
These interactions, which will be much stronger in machines with super-\Alfv{}ic particles, have the potential to drive AEs unstable, which can, in turn, redistribute the energetic particles (EPs), leading to radial transport. This transport would, in the best case, lead to reduced efficiency of the core plasma heating and could, in the worst case, lead to an ejection of energetic ions from the confined plasma. Such an ejection will add to the heat exhaust management problem of a reactor and could potentially lead to damage of unshielded plasma facing components.

The ITER tokamak, presently under construction, will be the first tokamak to have a significant number of fusion alpha particles present. Due to its size and parameters, the alpha particle population in ITER is predicted to have the potential to drive AEs unstable, however the amplitudes at which such modes saturate is important as it plays the determining factor in the levels of EP transport expected.

As the interactions between AEs and EPs take place over many spatio and temporal time scales, their modelling can be particularly difficult. In the past, various studies have investigated the issue of AEs in the 15$\ $MA scenario of ITER using a variety of different physical models. Previous to the work presented here, global gyrokinetic modelling of the scenario had only been performed using an eigenvalue code, and was limited to linear predictions~\cite{Lauber_ITER_2015, Pinches_ITER_2015}. In addition to this, hybrid MHD-kinetic simulations~\cite{Vlad2006, Todo2014}, perturbative hybrid gyrokinetic-drift kinetic analysis~\cite{Schneller2016} and perturbative hybrid MHD-drift kinetic~\cite{Gorelenkov2008, Rodrigues2015, Figueiredo2016, Fitzgerald2016, Isaev2017}. Local gyrokinetic analysis was also performed~\cite{Waltz2014}.
An interesting observation from previous work is that nonlinear predictions, performed with perturbative hybrid models, disagree about the likelihood of ITER seeing significant EP transport due to AEs, or how far from the threshold of significant EP transport~\cite{Chen2016} ITER may lie.

In this work, we use a more complete and consistent model available to try to answer some of these questions, namely the global nonlinear gyrokinetic model, by using the code ORB5~\cite{Lanti2020}. To do this, we have made several simplifications, which are outlined after a description of the scenario in \S\ref{sec:iter}.

	\section{Physical model}
In this work, the numerical results presented are obtained using the ORB5 code~\cite{Lanti2020}. This is a global electromagnetic gyrokinetic particle-in-cell (PIC) code which uses markers to sample the 5D phase space ($\vec{R}$, $v_\parallel$, $\mu$), with the equations independent of the gyrophase, and the magnetic moment ($\mu$) of a marker constant in the absence of collisions.

The perturbed part of the distribution function is then evolved according to the gyrokinetic Vlasov equation
\begin{equation}
        \der{t}{\delta f_s}
        =
        -\dot{\vec{R}} \cdot \left.\frac{\partial F_{0s}}{\partial \vec{R}}\right|_{\epsilon} -
        \dot{\epsilon}\frac{\partial F_{0s}}{\partial \epsilon}
\end{equation}
(where subscript $s$ denotes a plasma species $s$) according to the equations for the particle equations of motion
\begin{multline}
\label{eq:Rdot}
        \dot{\vec{R}} = \vpar \unitvecb - \vpar^2\frac{c m_\trm{s}}{q\bparstar}\vec{G} + \mu \frac{Bcm_\trm{s}}{q\bparstar}\unitvecb \times \frac{\grad{B}}{B} + \\
        \frac{\unitvecb}{\bparstar} \times \nabla \big\langle \phi - \vpar \Apar^{\trm{s}} - \vpar \Apar^{\trm{h}} \big\rangle - \frac{q_\trm{s}}{m_\trm{s}} \langle \Apar^{\trm{h}} \rangle \unitvecb^{*}
\end{multline}
\begin{multline}
\label{eq:vdot}
        \vpardot = \mu B \nabla \cdot \unitvecb + \mu \vpar \frac{c m_\trm{s}}{q_\trm{s} \bparstar}\vec{G}\cdot\grad{B} \
        - \mu \frac{\unitvecb \times \grad{B}}{\bparstar} \cdot \grad{\langle \Apar^{(s)} \rangle} \\
        -\frac{q_\trm{s}}{m_\trm{s}}\left[\unitvecb^{*}\cdot\nabla \big\langle \phi - \vpar \Apar^{(h)} \big\rangle + \pder{t}{}\langle\Apar^{(s)} \rangle \right]
\end{multline}
\begin{equation}
\label{eq:epsdot}
        \dot{\epsilon} = \vpar \vpardot + \mu \grad{B} \cdot \dot{\vec{R}}
\end{equation}

where
        $\vec{B} = \nabla \times \vec{A}$, $\unitvecb = \vec{B} / B$, $\bparstar = \unitvecb \times \vec{A}^{*}$
\begin{equation*}
        \vec{A}^{*} = \vec{A} + \left( \frac{m_\trm{s} c}{q_\trm{s}} \vpar + \langle \Apar^{(s)} \rangle \right) \unitvecb
\end{equation*}
\begin{equation*}
        \unitvecb^{*} = \frac{\nabla \times \vec{A}^{*}}{\bparstar} = \unitvecb - \left( \frac{c m_\trm{s}}{q \bparstar} \vpar + \langle \Apar^{(s)} \rangle \right) \vec{G}
\end{equation*}
\begin{equation*}
        \vec{G} = \unitvecb \times (\unitvecb \times (\nabla \times \unitvecb))
\end{equation*}
and with the gyroaveraged potential $\langle \phi \rangle = \oint \phi(\vec{R} + \vec{\rho})\df \alpha/(2\pi)$ with $\rho$ the gyroradius of the particle and $\alpha$ the gyro-phase.

These equations are coupled to the field equations, the gyrokinetic quasineutrality equation
\begin{equation}
        - \nabla \cdot \left[ \left( \sum_{s} \frac{q_\trm{s}^2 n_\trm{s}c^2}{T_\trm{s}} \rho_\trm{s}^2\right) \nabla_{\perp} \phi \right] = \sum_{s} q_\trm{s} n_\trm{1s}
\end{equation}
For the parallel Amp\`ere's law, the perturbed magnetic potential $\Apar$ is split in to the symplectic part $\Apar^{(s)}$, which is found from
\begin{equation}
        \label{eq:Ohm}
        \pder{t} \Apar^{(s)} + \unitvecb \cdot \grad{\phi} = 0
\end{equation}
and the Hamiltonian part $\Apar^{(h)}$, solved using the mixed-variable parallel Amp\`ere's law,
\begin{equation}
        \left(\sum_{s} \frac{\beta_\trm{s}}{\rho_\trm{s}^2} - \nabla_{\perp}^2 \right) \Aparh = \mun \sum_{s} j_{\parallel 1 s} + \nabla_{\perp}^2 \Apars
\end{equation}
with $n_\trm{1s} = \int \df{}^6 Z \delta f_\trm{s} \delta(\vec{R}+ \vec{\rho} - \vec{x})$ the perturbed gyrocenter density, $\rho_\trm{s} = \sqrt{m_\trm{s} T_\trm{s}}/(q_\trm{s} B)$ the thermal gyroradius, $q_\trm{s}$ the particle charge, $\df{}^6Z = \bparstar \df{} \vec{R}\df{} \vpar \df{} \mu \df{} \alpha$ the phase space volume, $j_\trm{1s} = \int \df{}^6 Z \vpar \delta f_\trm{s} \delta(\vec{R} + \vec{\rho} - \vec{x})$ the perturbed parallel gyrocenter current. Note that due to the scale separation between ion and electron Larmor radii, electrons are treated differently from ion species, and terms with $\rho_\trm{e}$ are neglected.

The particular method of using the mixed-variables formulation, and then applying a pullback procedure to the particles is described in reference~\cite{Mishchenko2019}, where its advantages for solving electromagnetic equations are discussed.

A set of straight field line coordinates are used, with radial coordinate $s=\sqrt{\psi/\psi_\trm{edge}}$ (where $\psi$ is the poloidal flux), toroidal angle $\varphi$, and poloidal angle
\[\chi = \frac{1}{q(s)} \int_0^\theta \frac{\vec{B}\cdot \grad\varphi}{\vec{B}\cdot \grad {\theta^\prime}} \df \theta^\prime\]
(where $\theta$ is the geometric poloidal angle).

The field equations are solved on a basis of cubic finite elements and using Fourier methods in the two angular directions. This allows the use of Fourier filtering, which helps to reduce noise by excluding non-physical modes which are far from being field aligned~\cite{McMillan2010}. Typically the local poloidal Fourier filter will retain modes with poloidal mode numbers $m_n(s) = \floor{n q(s)} \pm 5$ at each radial point for each toroidal mode n.

Linear simulations can be performed by linearizing equations for the particle trajectories and weights, equations~\ref{eq:Rdot},~\ref{eq:vdot},~and~\ref{eq:epsdot}.

	\section{Scenario description}
\label{sec:iter}
\begin{figure}
\includegraphics[width=0.25\textwidth]{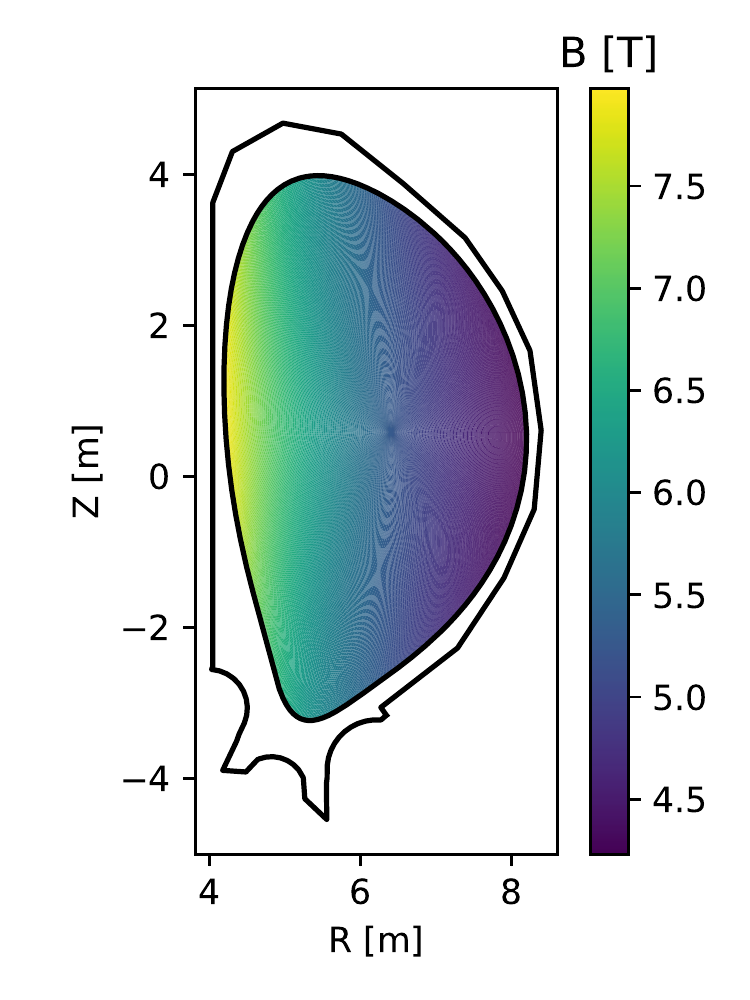}
\includegraphics[width=0.25\textwidth]{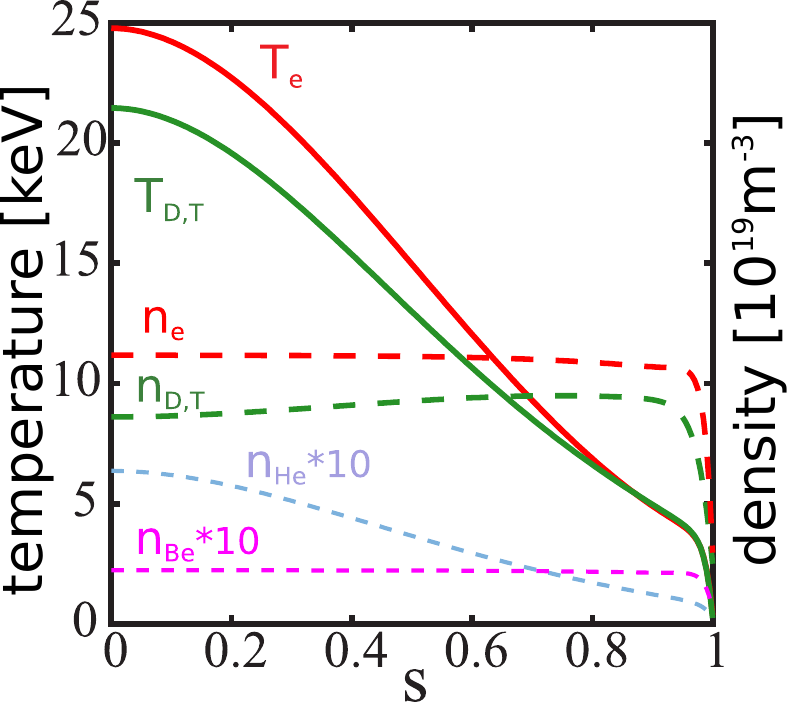}
\includegraphics[width=0.25\textwidth]{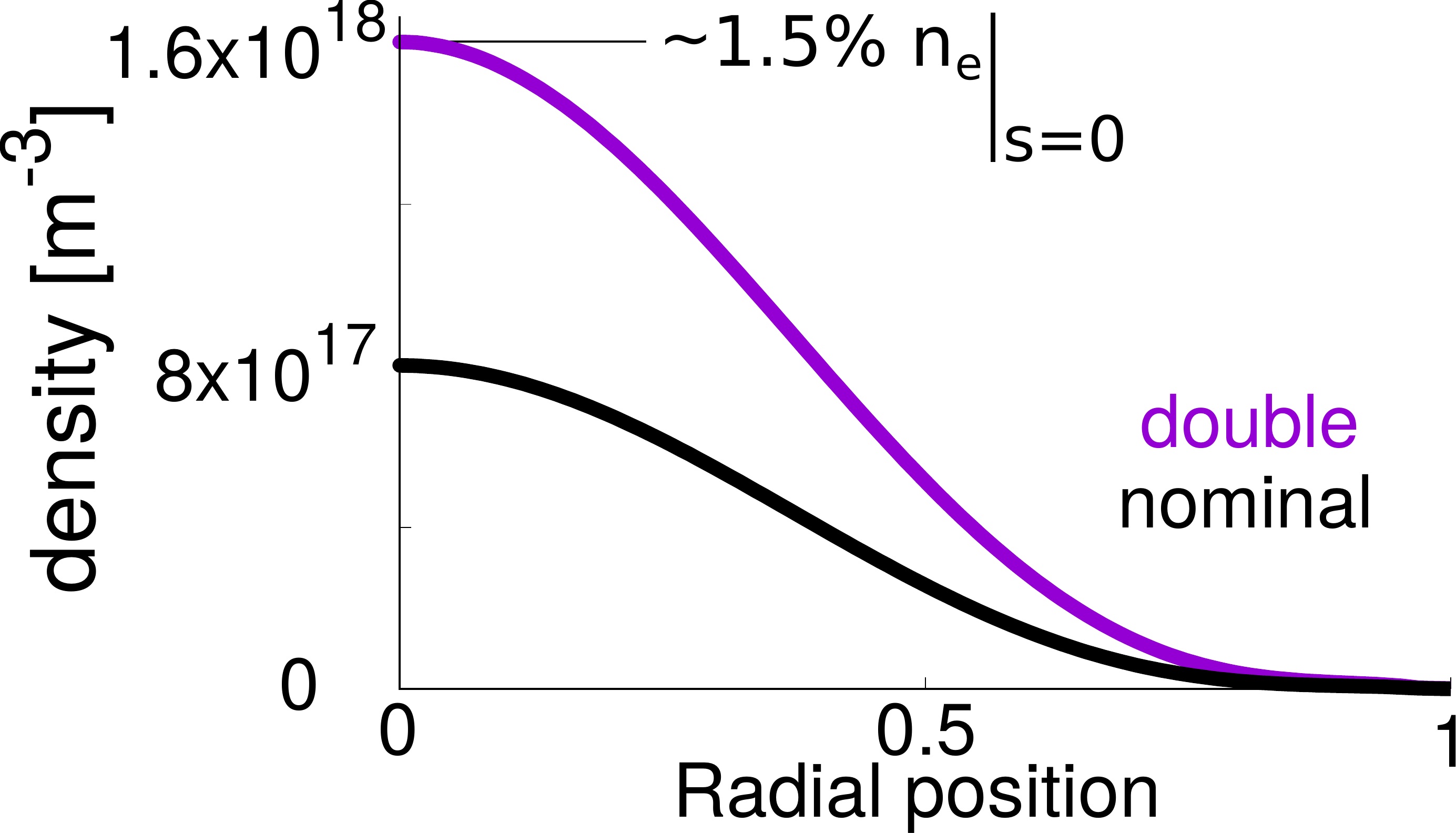}
\includegraphics[width=0.25\textwidth]{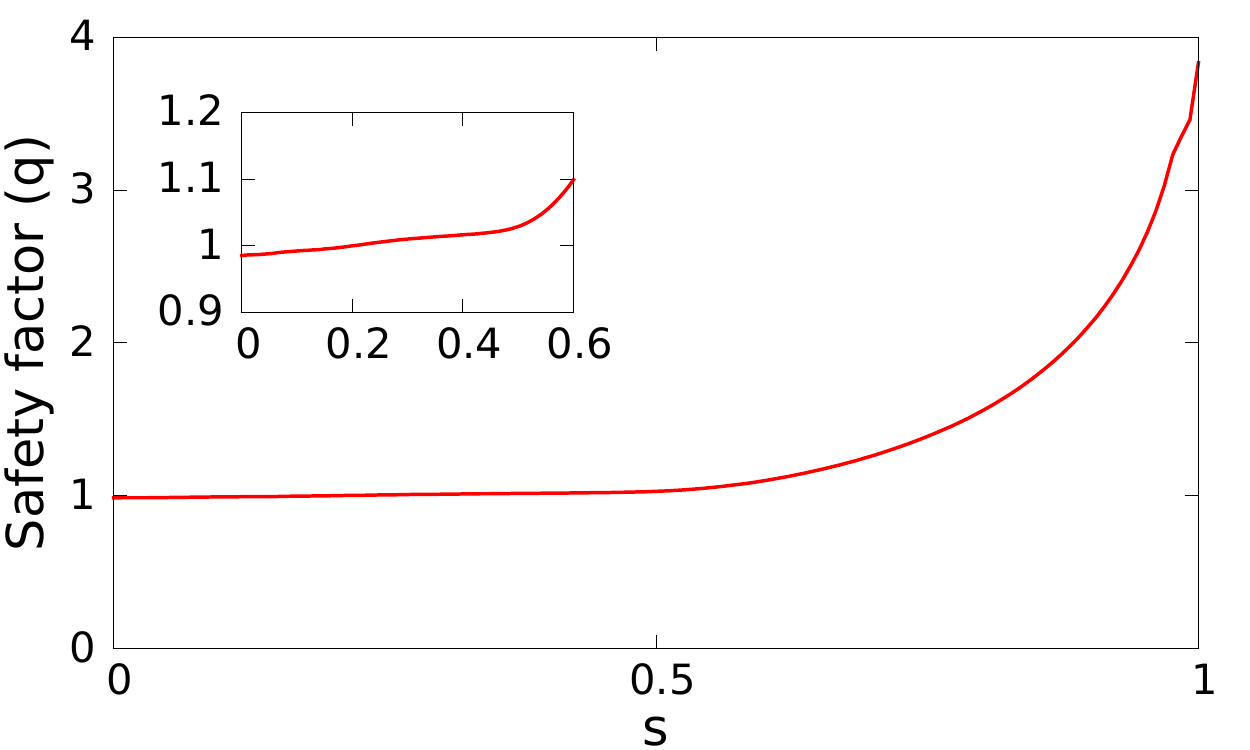}
\caption{(a) ITER 15MA scenario equilibrium; (b) density (dashed) and temperature (solid) profiles for the bulk species in the scenario; (c) the density profile of alpha particles for both nominal (black) and double (magenta) EP density; (d) the safety factor profile (with inset zoom). \\Figure (b) modified from reference~\cite{Lauber_ITER_2015}.}
\label{fig:iter_eq}
\end{figure}

Due to the burning D-T plasma and the significant population of alpha particles, the ITER~15MA baseline scenario~\cite{Polevoi2002} is a scenario of great interest for energetic particle physics.
The major radius, $R=6.2\ \trm{m}$; the minor radius, $a=2.0\ \trm{m}$; the magnetic field on axis $B_0=5.3\ \trm{T}$; the bulk ions are 50:50 deuterium and tritium, with some additional Helium ash and Beryllium impurities.
The electron temperature profile is peaked ($T_\trm{e,axis}/T_\trm{e,ped. top} \approx 5$), with the electron temperature on axis $T_{\trm{e},s=0} = 25\ \trm{keV}$.
The ion temperature profile has a similar shape to the electron profile, but is slightly less hot, with $T_{\trm{i},s=0} = 21\ \trm{keV}$.
The density profiles of the ions and electrons (linked via quasineutrality and the impurity density profiles) are very flat, with the electron density profile marginally peaked and the ion temperature profile marginally hollow.
The on-axis electron density $n_{\trm{e},s=0} \approx 1.1 \cdot 10^{20}\ \trm{m}^{-3}$. The temperature and density profiles for the background species are shown in figure~\ref{fig:iter_eq}(b).

The energetic particles, a population of fusion alpha particles with birth energy $3.5\ \trm{MeV}$ are peaked in the core of the plasma where $n_{\trm{EP},s=0}/n_{\trm{e},s=0} \approx 0.75 \%$.
The energetic particle density profile is shown in figure~\ref{fig:iter_eq}(c).

The equilibrium was considered by taking the nominal equilibrium file and running the CHEASE MHD equilibrium code~\cite{CHEASE}.

\subsection{Scenario simplifications}
Throughout this work, we shall make various modifications to the nominal scenario outlined above.
We avoid the difficulty in treating the gradients of the pedestal by setting the gradients of the bulk \{ion, electron\} \{density,temperature\} profiles to zero at $s=0.91$.
We additionally consider one isotope of mass $^{2.5}\trm{DT}$ instead of a mixture of 50:50 D:T for the bulk ions.
We neglect the effect of the cold impurities (He, Be) and therefore take $n_\trm{DT}(s)=n_\trm{e}(s)$.
We consider a double density EP profile.
We treat the EPs drift-kinetically rather than gyrokinetically.
We approximate the distribution of EPs, nominally an isotropic slowing down distribution function with birth energy $3.5\ \trm{MeV}$, with a Maxwell-Boltzmann distribution with temperature $900\ \trm{keV}$.
We note that the isotope and impurity effects have been previously investigated in~\cite{Lauber_ITER_2015}.

Throughout this work, when modifying the plasma profiles we subsequently set $n_\trm{e}$ so as to preserve quasineutrality (QN). Given that $n_\trm{EP} \ll n_\trm{e}$, we do not anticipate any significant difference between changing $n_\trm{e}$ and $n_\trm{DT}$ to preserve QN when varying $n_\trm{EP}$.

	\section{Numerical parameters}
\label{sec:numerical}
When performing numerical simulations, we wish to balance physical fidelity, numerical accuracy and cost. We test the numerical properties of the simulations with one of the more realistic linear simulations using the same simulation setup as shown later in \S\ref{sec:linear_stability}.

Given that the time taken to perform a single time step for a single-n annular simulation in this study is approximately 0.5 core-hours (with full radius simulations taking more time, and nonlinear multi-mode simulations scaling approximately with the number of modes), we wish to reduce the cost as well as the time taken to achieve results.
\subsection{Convergence of growth rate with timestep and electron mass}
In simulations with kinetic electrons, the mass of the electrons simulated affects the size of the time step that we are allowed to take, however below a certain threshold, the effect of the electron mass on the physical result is expected to be minimal.
\begin{figure}
	\centering
        \includegraphics[width=0.49\textwidth]{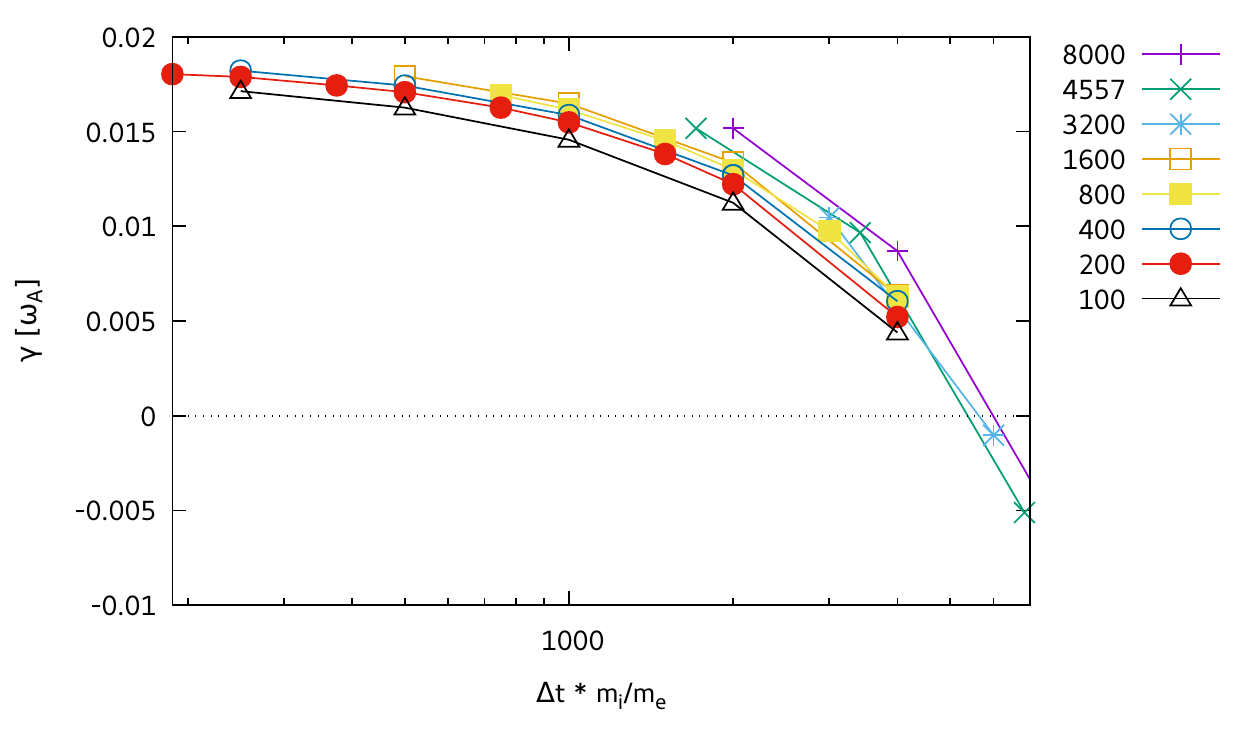}
        \includegraphics[width=0.49\textwidth]{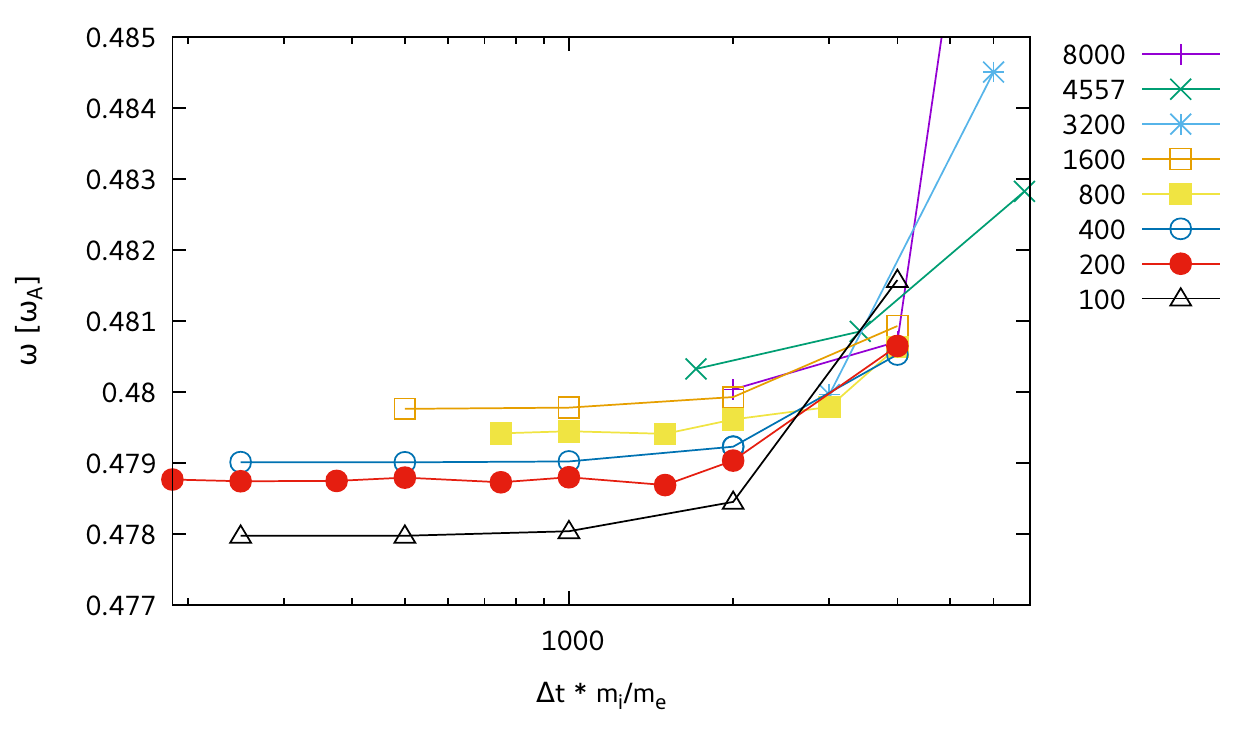}
	\caption{Growth rate (a) and frequency (b) on log-linear scales taken from simulations of an $n=26$ TAE run with shaped profiles on an annular ($0.2 \le s \le 0.7$) domain for different values of the electron mass and numerical time step. The legend refers to the ratio between ion and electron mass, which also normalizes the values of the horizontal axis.}
	\label{fig:mr_dt}
\end{figure}
A study of the convergence of the growth rate with respect to the time step, at a number of different values for the electron mass is shown in figure~\ref{fig:mr_dt}, and we can draw a few conclusions.
Firstly, the difference in the physical properties, namely the damping and frequency vary only very weakly with varying the ion/electron mass ratio, once the ratio is larger than 100 (we observe the difference in $\gamma$ between mass ratios of 200 and realistic (4557) to be $\le 20\%$).
Secondly, the individual curves of convergence of the measured growth rate are close to overlapping once scaled by the electron mass. Whilst in this figure, as it is difficult to completely decouple a vertical shift (due to changed electron damping) from a horizontal shift (scanning the simulation time step), it is clear that the scaling of the time step convergence with respect to the mass is much closer to the inverse of the mass than the inverse of the square root of the mass as one might na\"ively assume.
Given the observed behaviour of the time step and mass ratio, we proceed in sections \S\ref{sec:linear_stability} and \S\ref{sec:nonlinear} with $m_i/m_e=200$ and $dt=1.875$ (the 3rd red point from the left of figure~\ref{fig:mr_dt}).

\begin{figure}
	\centering
        \includegraphics[width=0.4\textwidth]{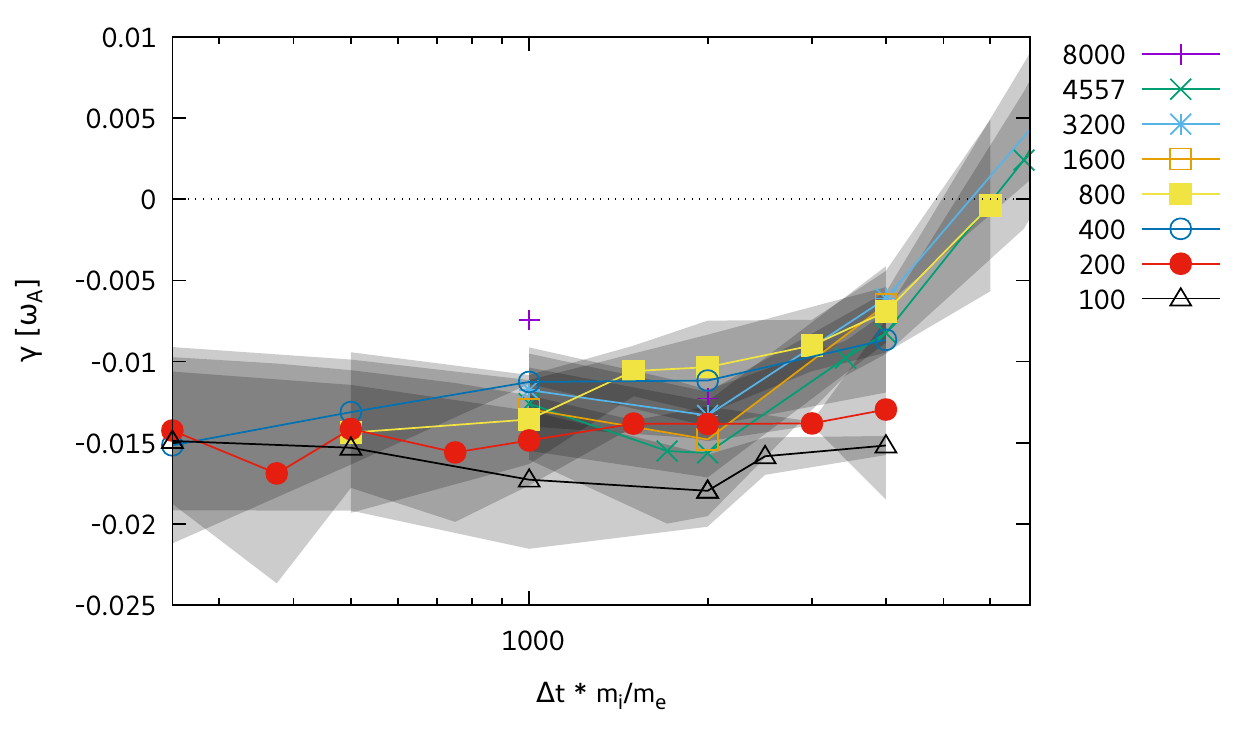}
	\caption{The uncertainty on the fit of the growth rates, as per figure~\ref{fig:mr_dt}, approximating the uncertainty by taking the discrepancy between measuring the growth rate by fitting the entire signal or by fitting only the extrema of the signal. With zero alpha particle density.}
        \label{fig:mr0_dt_3_error}
	\label{fig:mr_dt_error}
\end{figure}

In figure~\ref{fig:mr0_dt_3_error}, we show the decay rate measured in the absence of energetic particles.
The grey bands mark the approximate error bars, as calculated by comparing the fit of the decay of the signal measured with two different methods, fitting only the maxima and fitting the whole signal.
It should also be said that the time range over which such a fit is made affects the measured decay rate, with the initial phase decaying more steeply than the later phase.
As the amplitude of the signal decreases over time, thus making long decaying simulations noisy, (for this reason, the fitting for growing simulations in figure~\ref{fig:mr_dt} was done between $t=\{20000,40000\}$, and for figure~\ref{fig:mr0_dt_3_error} between $t=\{0,20000\}$, which makes an absolute comparison somewhat difficult.
However we observe that the fitted decay rate is largely in the range $\gamma = -(0.01$--$0.015)\ \omega_{A}$ ($\approx -(2$--$3)\%$), acknowledging that this is likely an over estimate of the observed damping\footnote{As previously mentioned, in the damped simulations, we fit an earlier time window where we see a steeper decay (we see also the decay of the initial perturbation which decays faster than the true decaying eigenmode).}.

\begin{figure}
	\centering
	\includegraphics[width=0.4\textwidth]{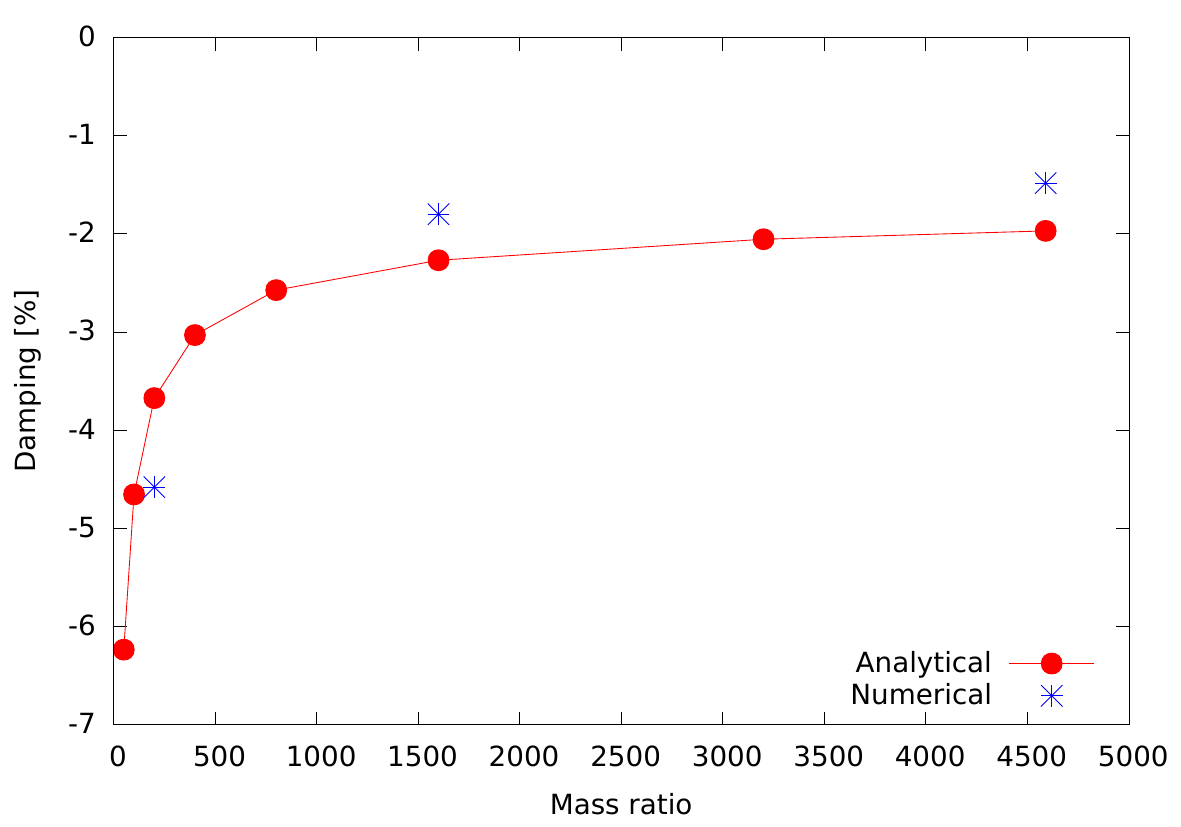}
	\caption{Damping rates as a function of $m_i/m_e$, calculated with LIGKA, measured in units of the mode real frequency. Data range from realistic mass ratios (4557) down to heavy electrons (50). The red points, labelled `analytical' were calculated using the analytical version of LIGKA, whereas the blue points used the full numerical orbits for integrations.}
	\label{fig:ligka_electron_mass}
\end{figure}

An additional comparison is to compare the results obtained with artificially heavy electrons with the results for damping obtained with the linear gyrokinetic eigenvalue solver LIGKA~\cite{ligka}, which was modified during this study to run with non-realistic electron mass.
That model retains the changes to the damping due to the change of the mass of electrons, but the numerical cost of the solution and quality of the solution should be independent of the electron mass. Figure~\ref{fig:ligka_electron_mass} shows the dependence of the damping on the electron mass, calculated both with analytical coefficients as well as the more accurate model with numerically calculated coefficients.
We observe here that the realistic mass ratio shows damping rate in the range of $1.5$--$2$\%.
In comparison with that measured from ORB5, this is slightly weaker damping and we see a steeper increase in damping with lower mass ratios.

Overall, we see good agreement between ORB5 and LIGKA for relevant values of the mass ratio. For very small mass ratios, the differences are not yet understood, but are of lesser interest.

	\section{Linear modelling}
\label{sec:linear}
\subsection{Initial simulations of TAEs}
We start by demonstrating \Alfv{} eigenmodes in our simplest approximation of the scenario. In this case, we have flattened the temperature and density profiles of the background species for both ions and electrons, fixing them at approximately their values at mid-radius. This case also has a very slightly scaled safety factor, such that $q_\trm{axis} = 0.99$.

\begin{figure}
        \centering
        \includegraphics[width=0.23\textwidth]{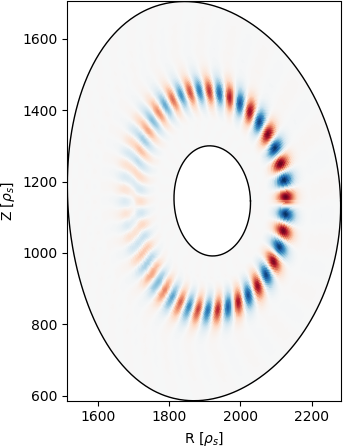}
        \includegraphics[width=0.23\textwidth]{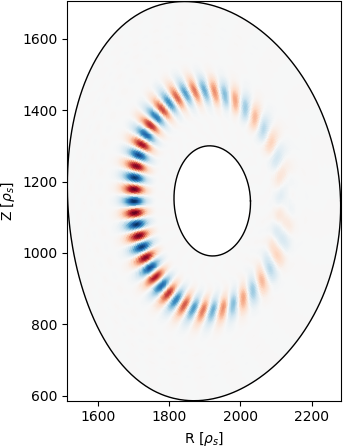}
        \caption{(a) Electrostatic and (b) electromagnetic potential perturbations in the poloidal plane, showing $n=24$ TAE simulated on an annulus with reduced mass ratio.}
        \label{fig:n24_lin_flat_annulus_2d_es_and_em}
\end{figure}

\begin{figure}
	\centering
	\includegraphics[width=0.45\textwidth]{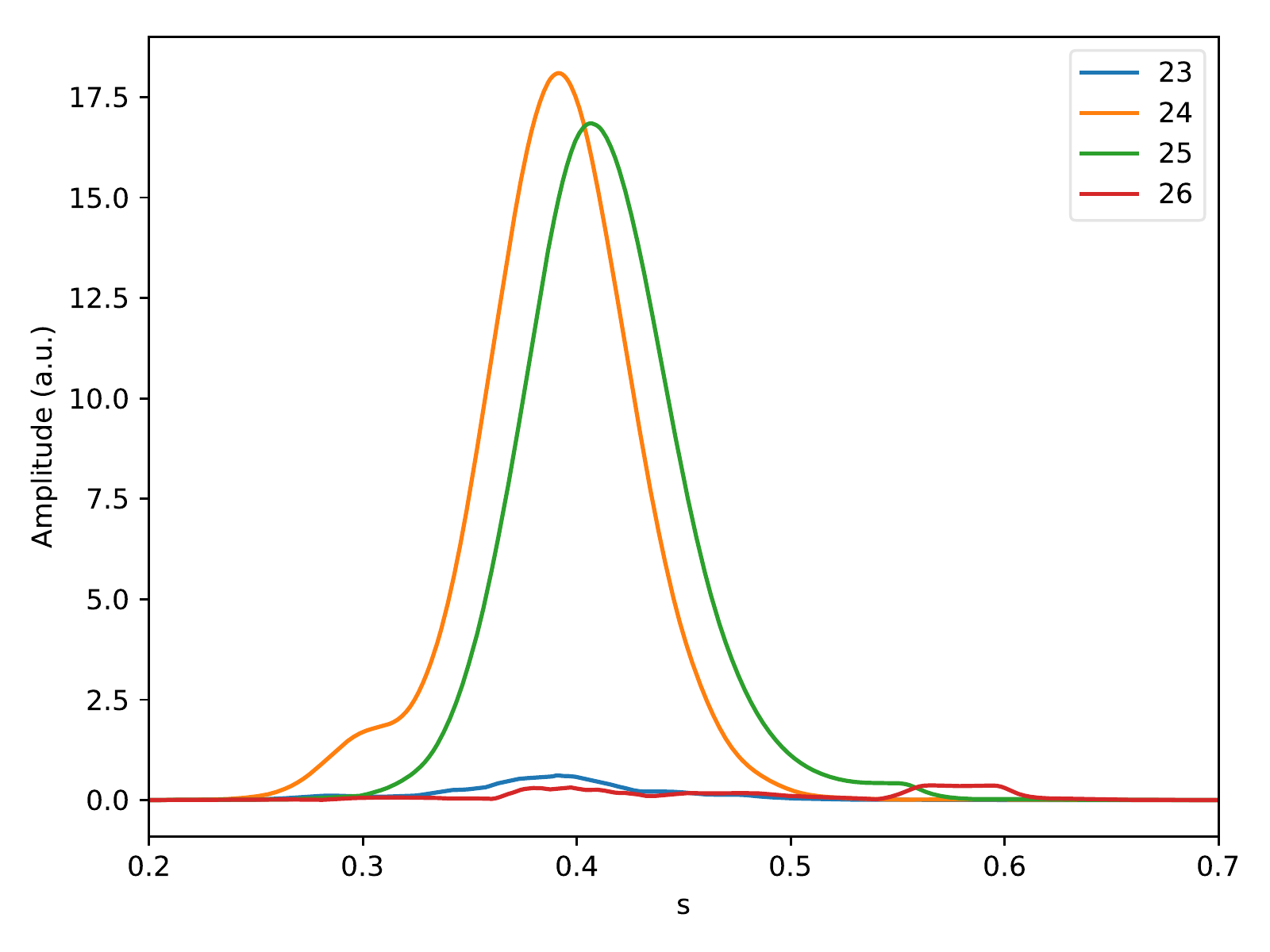}
	\caption{Mode structure for an $n=24$ TAE simulated with flat background profiles, double the nominal EP density, reduced mass ratio of $m_\trm{i}/m_\trm{e}=200$, simulated on an annulus of $s=0.2$--$0.7$. The labels refer to different poloidal harmonics of the electrostatic potential.}
	\label{fig:n24_lin_flat_annulus}
\end{figure}

\begin{figure}
	\centering
	\includegraphics[width=0.45\textwidth]{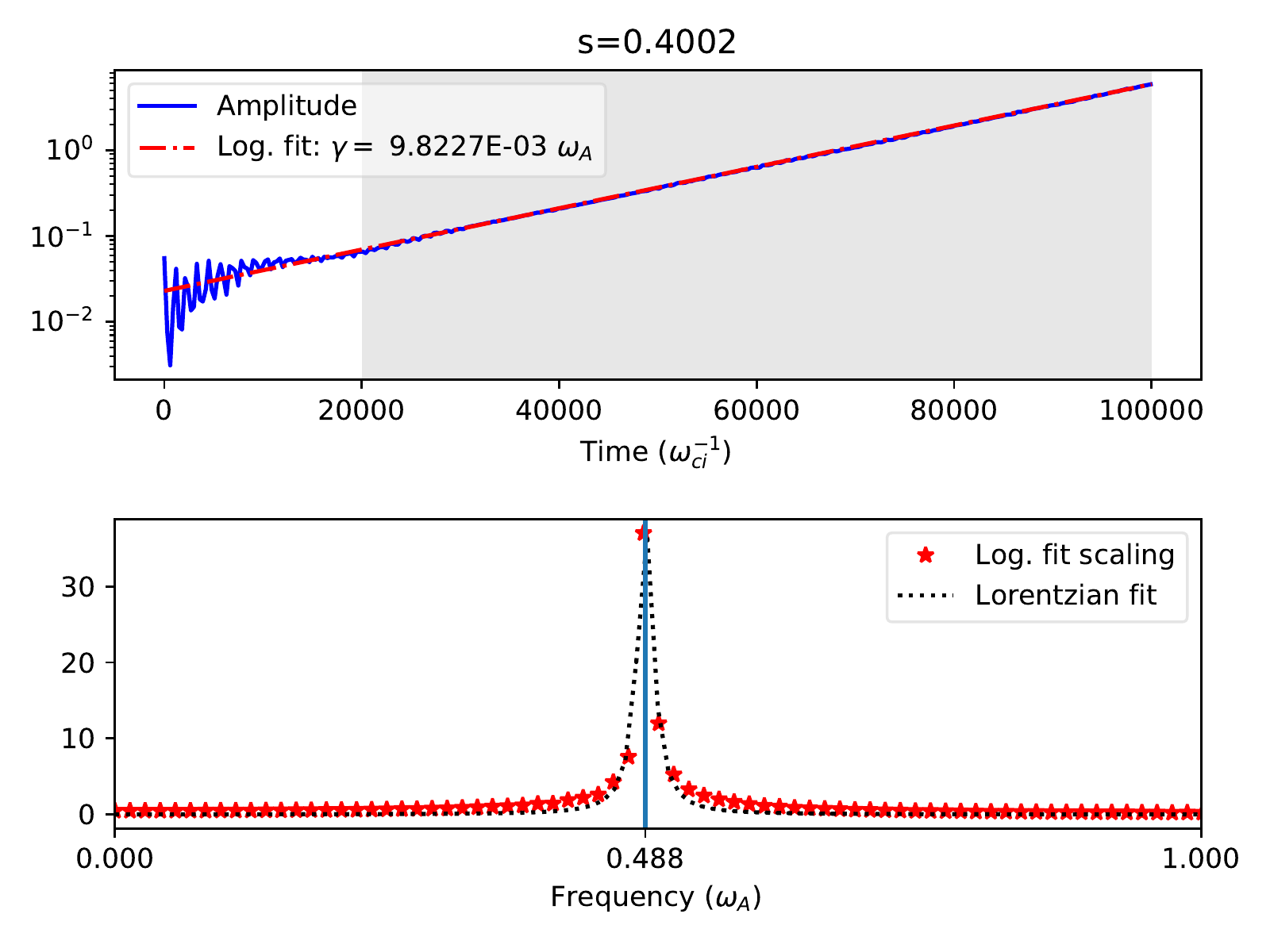}
	\caption{Frequency and growth rate analysis performed for the same $n=24$ simulation as shown in figure~\ref{fig:n24_lin_flat_annulus}. In the upper part of the figure, we show the envelope evolution and the fit for the growth rate, which is written in the label of the figure. In the lower part, we plot the amplitude of each frequency bin (in units of the \Alfv{} frequency) of the normalized signal after applying a Fourier transform. For the purpose of interpolating the frequency, we also fit a Lorentzian function to the values in the bins of the Fourier transform.}
	\label{fig:n24_lin_flat_annulus_freq}
\end{figure}

\begin{figure}
	\centering
        \includegraphics[width=0.23\textwidth]{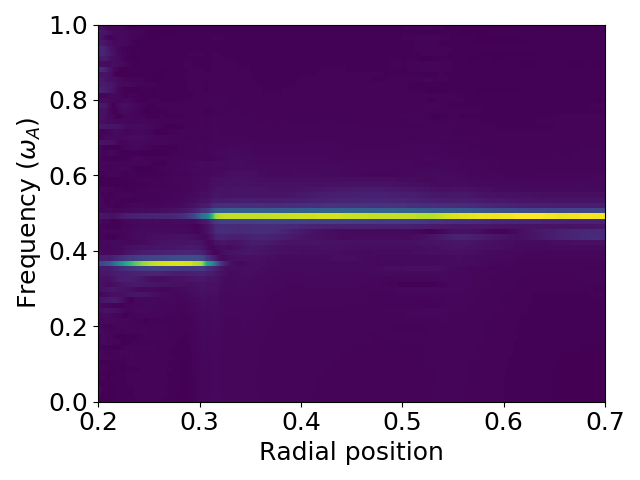}
        \includegraphics[width=0.23\textwidth]{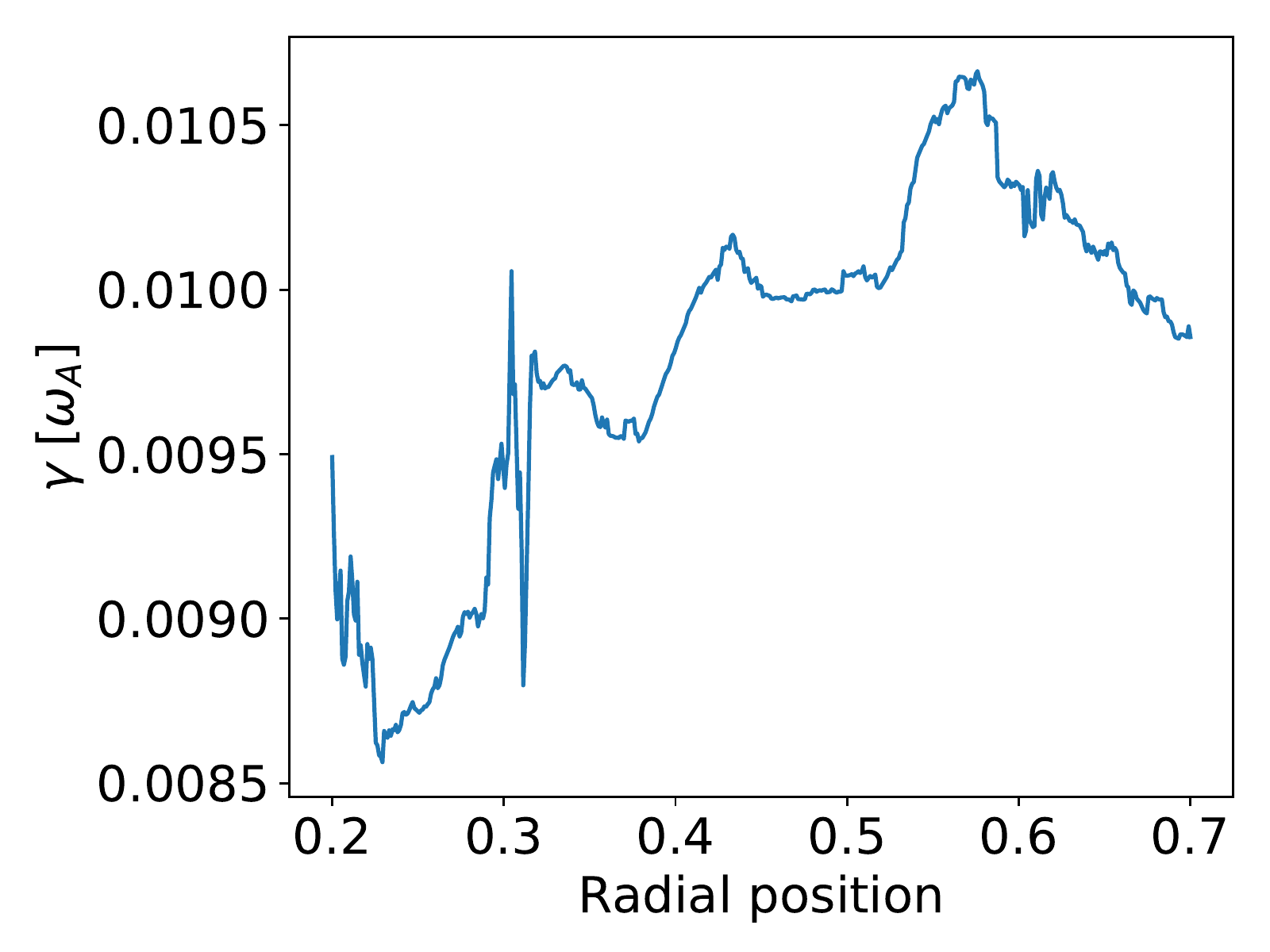}
	\caption{Measurements of the frequency and growth rate at different radial positions, applying the same procedures as in figure~\ref{fig:n24_lin_flat_annulus_freq}, which equates to a cut of these figures at radial position $s=0.4$. On the left, we plot the frequency spectrum, noting that each radial position is normalized independently based on its fit in the right figure.}
	\label{fig:n24_lin_flat_annulus_freq_radial}
\end{figure}

On an annulus (radial width $s=\{0.2,0.7\}$), we perform linear simulations of an $n=24$ TAE. Figure~\ref{fig:n24_lin_flat_annulus_2d_es_and_em} shows a slice of the poloidal plane showing (left) the electrostatic potential ($\phi$) and (right) the parallel electromagnetic potential ($A_{\parallel}$).
We can observe the characteristic properties of these high-n TAEs, a radially well localized ballooning structure in $\phi$ and the anti-ballooning structure of $A_\parallel$, both localized around the position of the $q=24.5/24$ surface. In figure~\ref{fig:n24_lin_flat_annulus}, we show the radial profile of the poloidal Fourier decomposition of this signal, and in figure~\ref{fig:n24_lin_flat_annulus_freq} the frequency and growth rate analysis used in this work. In this case, we see a frequency, $\omega = 0.488\ \wa$ and a growth rate, $\gamma = 9.8\cdot 10^{-3}\ \wa$, or 2\% of $\omega$, with figure~\ref{fig:n24_lin_flat_annulus_freq_radial} showing that this frequency is dominant at radial positions $s > 0.3$, with only small variation in the fitted growth rate across the domain.

\begin{figure}
	\centering
	\includegraphics[width=0.45\textwidth]{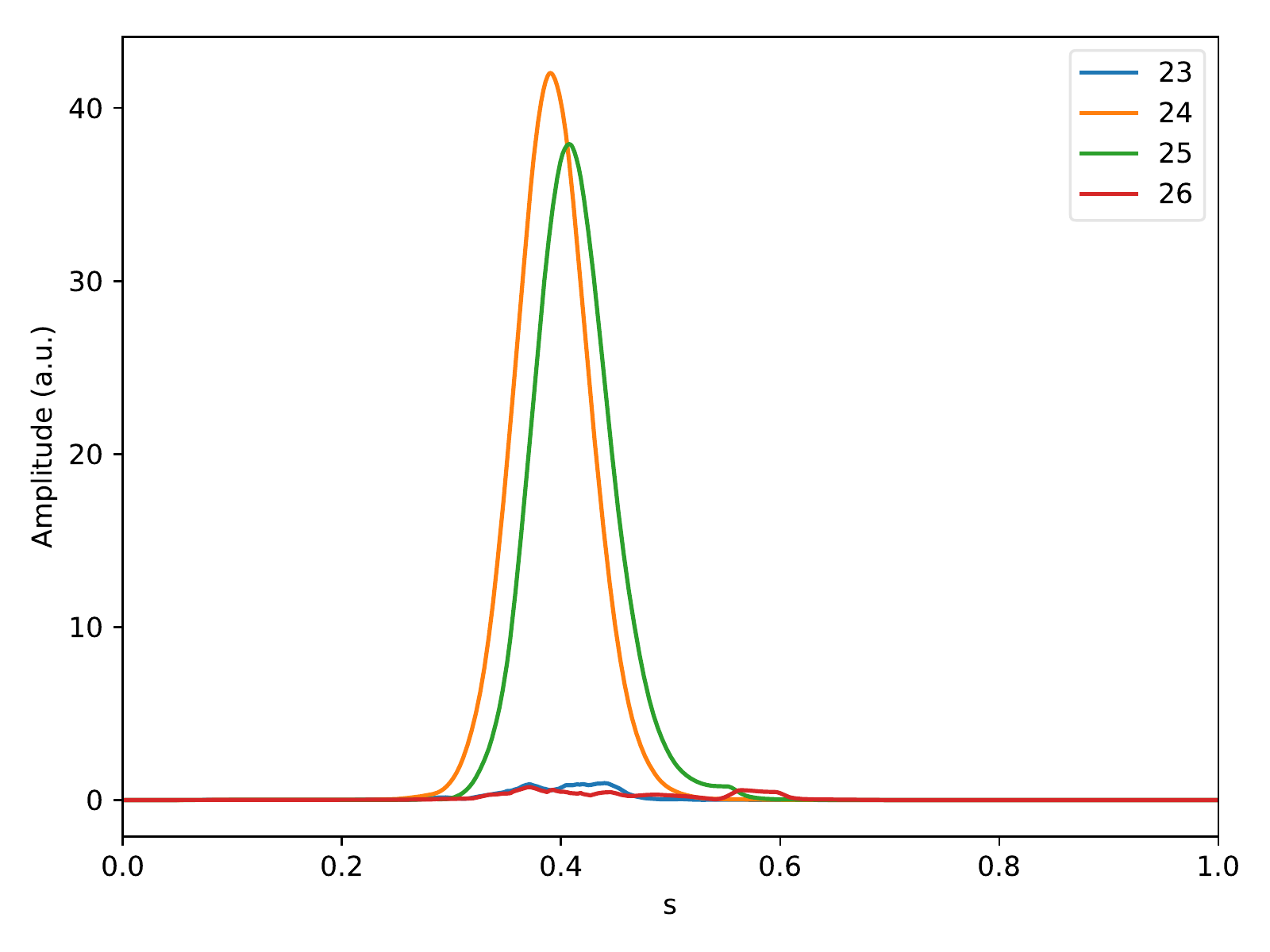}
	\caption{Mode structure for an $n=24$ TAE simulated with flat background profiles, double the nominal EP density, reduced mass ratio of $m_\trm{i}/m_\trm{e}=200$, simulated on the entire radial domain. The labels refer to different poloidal harmonics of the electrostatic potential.}
	\label{fig:n24_lin_flat_full}
\end{figure}

Moving to a global radial domain, $0.0 < s < 1.0$, we repeat the same case, shown in figure~\ref{fig:n24_lin_flat_full}, observing the same mode structure, frequency, and growth rate.


\subsection{Other \Alfv{} eigenmodes}
\begin{figure}
        \centering
        \includegraphics[width=0.39\textwidth]{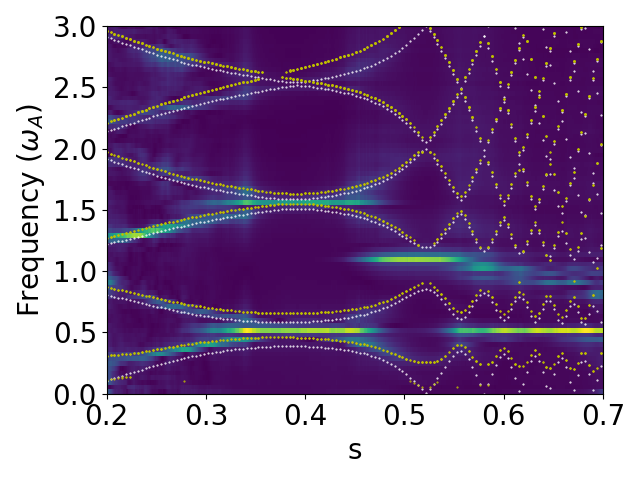}
	\caption{Radial frequency spectrum for $n=24$ simulated with zero EP density. We overplot the \Alfv{} continuum from the ideal MHD version of LIGKA in white and the \Alfv{} continuum from the analytical kinetic version of LIGKA in yellow.}
        \label{fig:n24_lin_flat_annulus_freq_spectrogram_ep0_initg}
	\label{fig:n24_lin_flat_annulus_freq_spectrogram_ep0}
\end{figure}

\begin{figure}
	\centering
	\includegraphics[width=0.49\textwidth]{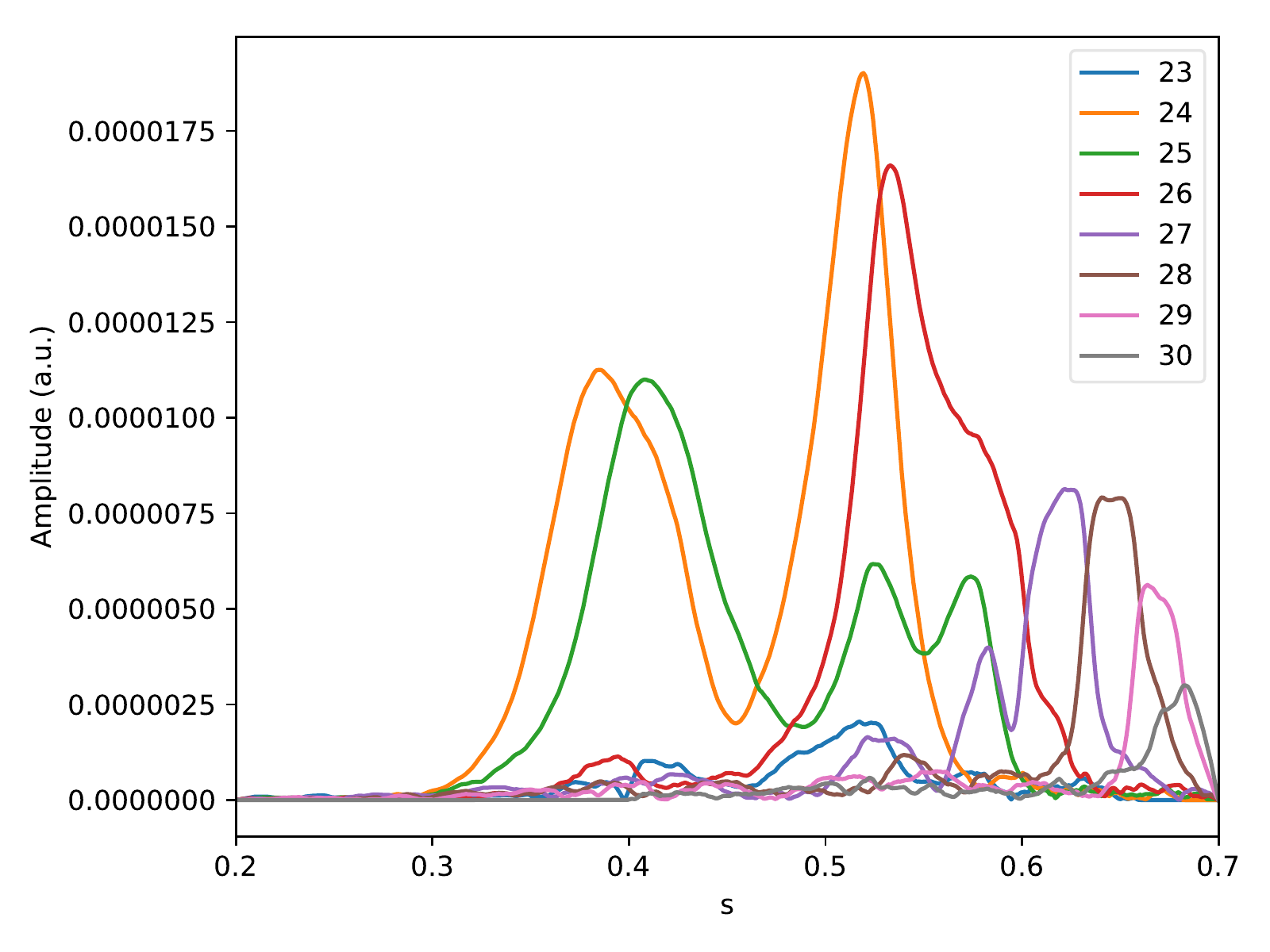}
	\caption{Radial harmonic structure for $n=24$ simulation without EPs present. The simulation corresponds to figure~\ref{fig:n24_lin_flat_annulus_freq_spectrogram_ep0_initg}.}
	\label{fig:n24_lin_flat_annulus_ep0}
\end{figure}

In addition to the TAEs, which we previously observed to be strongly driven, we are also interested in studying any other \Alfv{} eigenmodes which may be present in ITER.
If we perform a simulation with a strongly driven mode, or a decaying simulation where we initialize a perturbation close to one of the eigenmodes of the system, then our analysis (of mode structure, frequency, damping) will show this mode.
However, it is also possible to study weakly damped modes, or to assess the damping/drive of all (or at least more) modes in the system by choosing a more general initial perturbation.
This we do in figures~\ref{fig:n24_lin_flat_annulus_freq_spectrogram_ep0}~and~\ref{fig:n24_lin_flat_annulus_ep0}.
Here we initialize a radially broad, $n=24$, approximately field aligned density perturbation in the absence of EPs, and analyse the decaying simulation.
Figure~\ref{fig:n24_lin_flat_annulus_freq_spectrogram_ep0} shows that there are several distinct modes present in the simulation.
A TAE (and a weaker, higher frequency, NAE) at $s\approx0.4$, an EAE at $s\approx0.52$, and an edge TAE at $s>0.6$.
We also see evidence of additional, less distinct, EAEs further out, as well as direct measurements of the \Alfv{} continuum itself.
In figure~\ref{fig:n24_lin_flat_annulus_ep0}, we see the corresponding mode structures, with the $m=\{24,26\}$ EAE at $s=0.52$ dominant.

\begin{figure}
        \centering
        \includegraphics[width=0.4\textwidth]{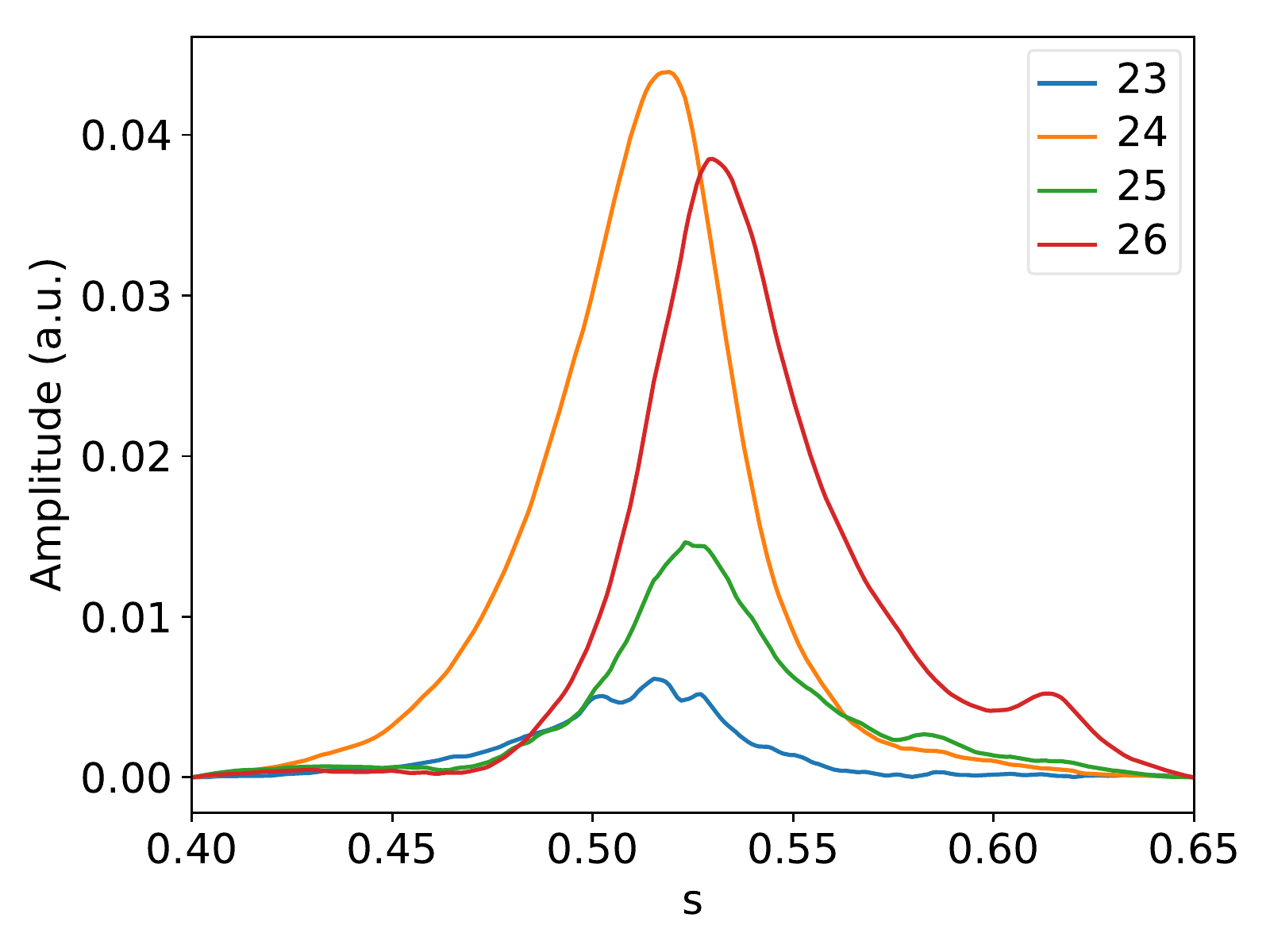}
	\caption{Radial harmonic structure for an $n=24$ EAE simulation without EPs present.}
        \label{fig:n24_eae}
\end{figure}

To further study the EAE, we isolate this mode by initializing an $n=24$, $m=\{24,26\}$ perturbation centred at $s=0.52$, on a reduced radial annulus of $0.4 < s < 0.65$.
The final mode structure (at $t=40000\ \wci^{-1}$) is shown in figure~\ref{fig:n24_eae} with the measured frequency and damping rate $\omega = 1.091\ \wa$, $\gamma/\omega = -0.58\%$.

\begin{figure}
	\centering
        \includegraphics[width=0.4\textwidth]{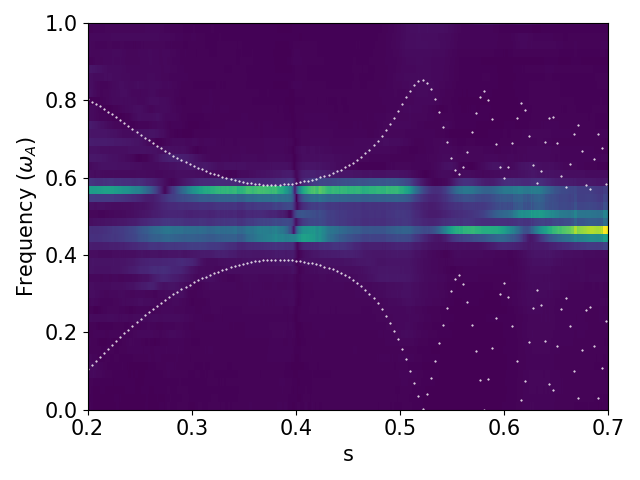}
        \label{fig:n24_oddtae_ep0_freq_spectrogram}
	\caption{Analysis of the frequency for $n=24$ TAEs without energetic particles, initialized with odd parity. The \Alfv{} continuum, as calculated from the ideal MHD version of LIGKA, is overplotted in white.}
	\label{fig:n24_oddtae_ep0_freq}
\end{figure}

Also of interest, and also expected to be weakly damped, are the odd-parity TAEs~\cite{Mett1992,Breizman1995}, whose frequency sits close to the upper continuum accumulation point.
Because of the different coupling of the harmonics, the odd TAEs have an anti-ballooning structure in $\phi$ and a ballooning structure in $A_\parallel$.
Therefore, to study odd TAEs, we initialize a pertubation with $n=24$, $m=\{24,25\}$ at the position of the TAE, but where there relative sign of the amplitude of the $m=25$ is flipped with respect to the $m=24$ harmonic.
The frequency analysis of this is shown in figure~\ref{fig:n24_oddtae_ep0_freq}, with the peaks at $s=0.35$ fit at $\omega_1=0.4478$ and $\omega_2=0.5547\ \wa$, with $\omega_2$ matching the frequency of the upper continuum accumulation point as overlaid, calculated with LIGKA.

\subsection{Linear mode stability}
\label{sec:linear_stability}
Having demonstrated proof of principle TAEs and other AEs in the previous section, we now proceed with an assessment of the linear stability of TAEs in the shaped profile equilibrium with $q_0=0.985$.
We consider a broad range of modes, from $n=10$ up to $n=40$.
As with the previous results, these are considered with a Maxwellian distribution function with double the nominal density for the EPs and neglecting the gyroaveraging operator on the EPs unless otherwise specified.

\begin{figure}
	\centering
        \includegraphics[width=0.49\textwidth]{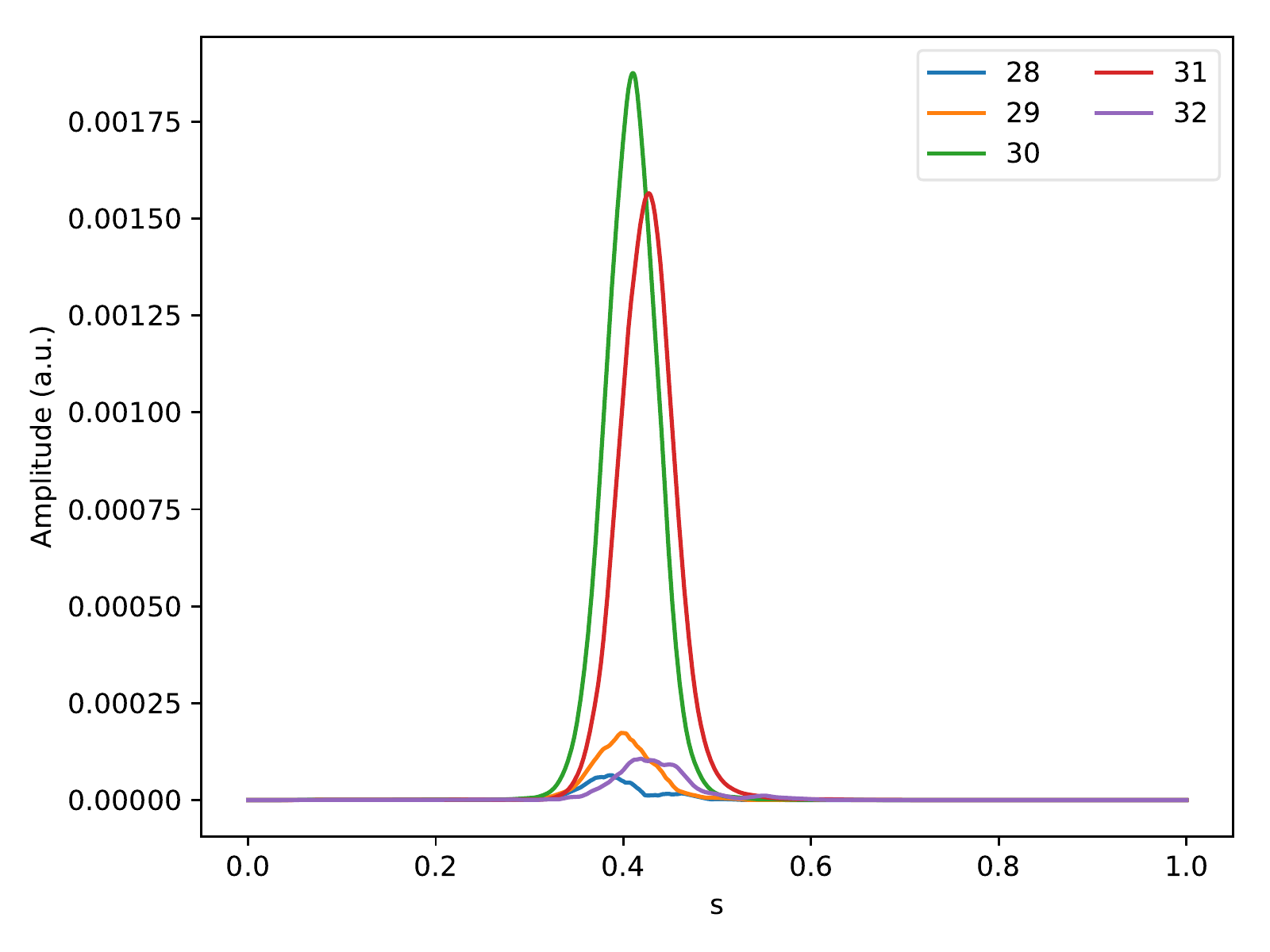}
	\caption{Mode structure of an $n=30$ TAE.}
        \label{fig:n30_global_radial}
\end{figure}

\begin{figure}
\centering
        \includegraphics[width=0.45\textwidth]{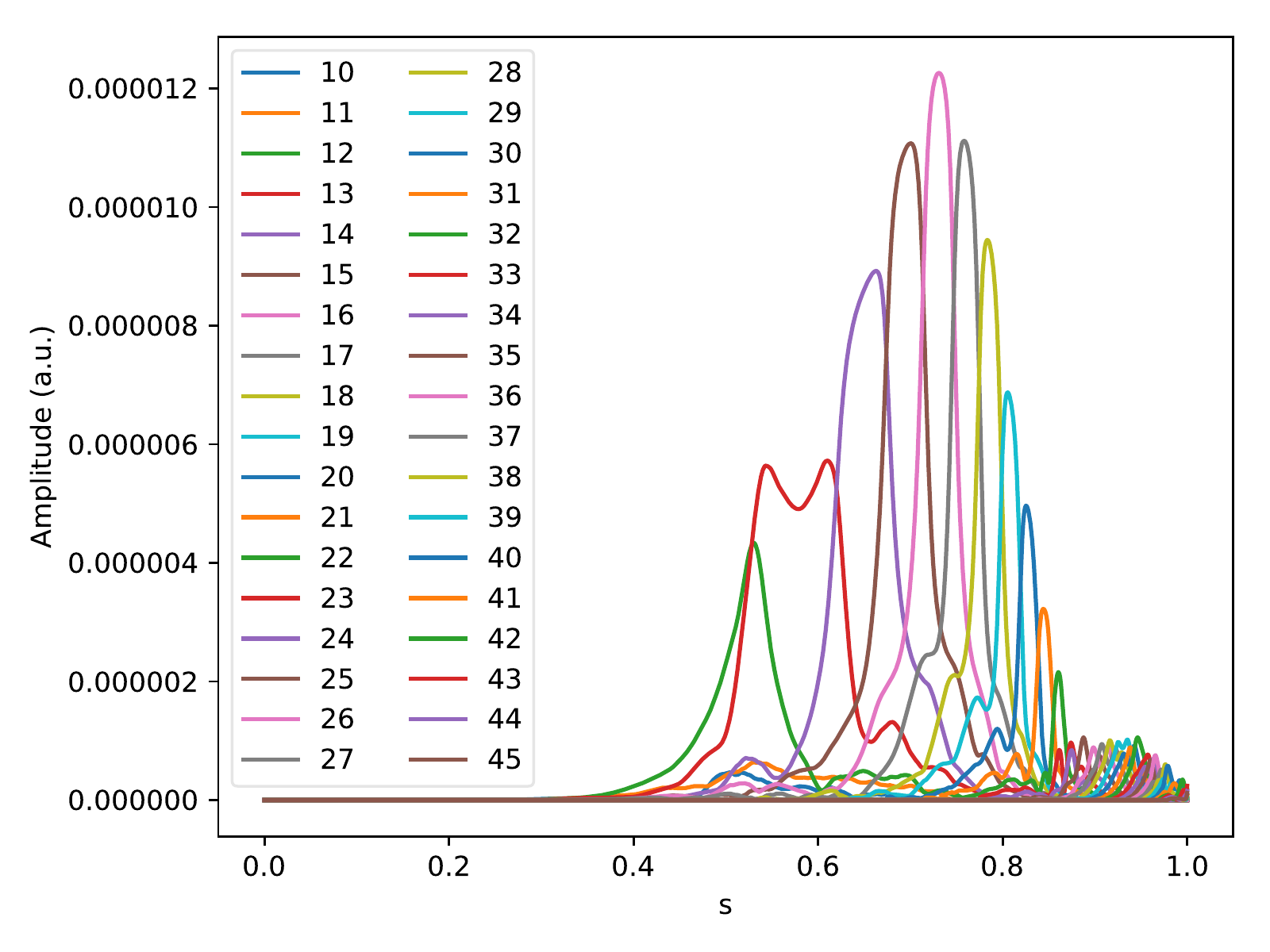}
        \includegraphics[width=0.45\textwidth]{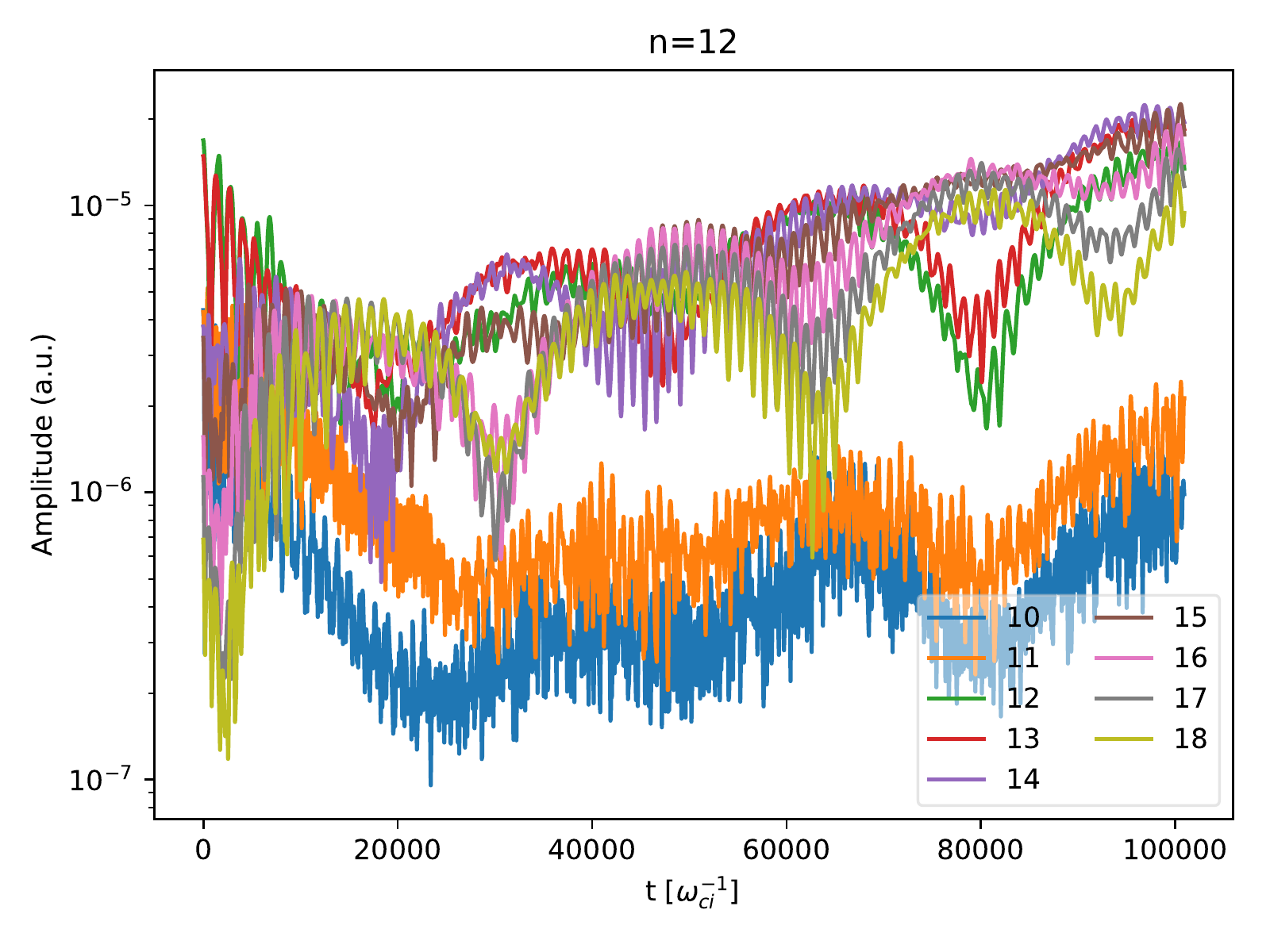}
        \caption{$n=12$ TAEs simulated over a global domain. (a) radial structure of poloidal harmonics, (b) temporal evolution of peak values for each poloidal harmonic.}
        \label{fig:n12_global}
\end{figure}

\begin{figure}
	\centering
	\includegraphics[width=0.49\textwidth]{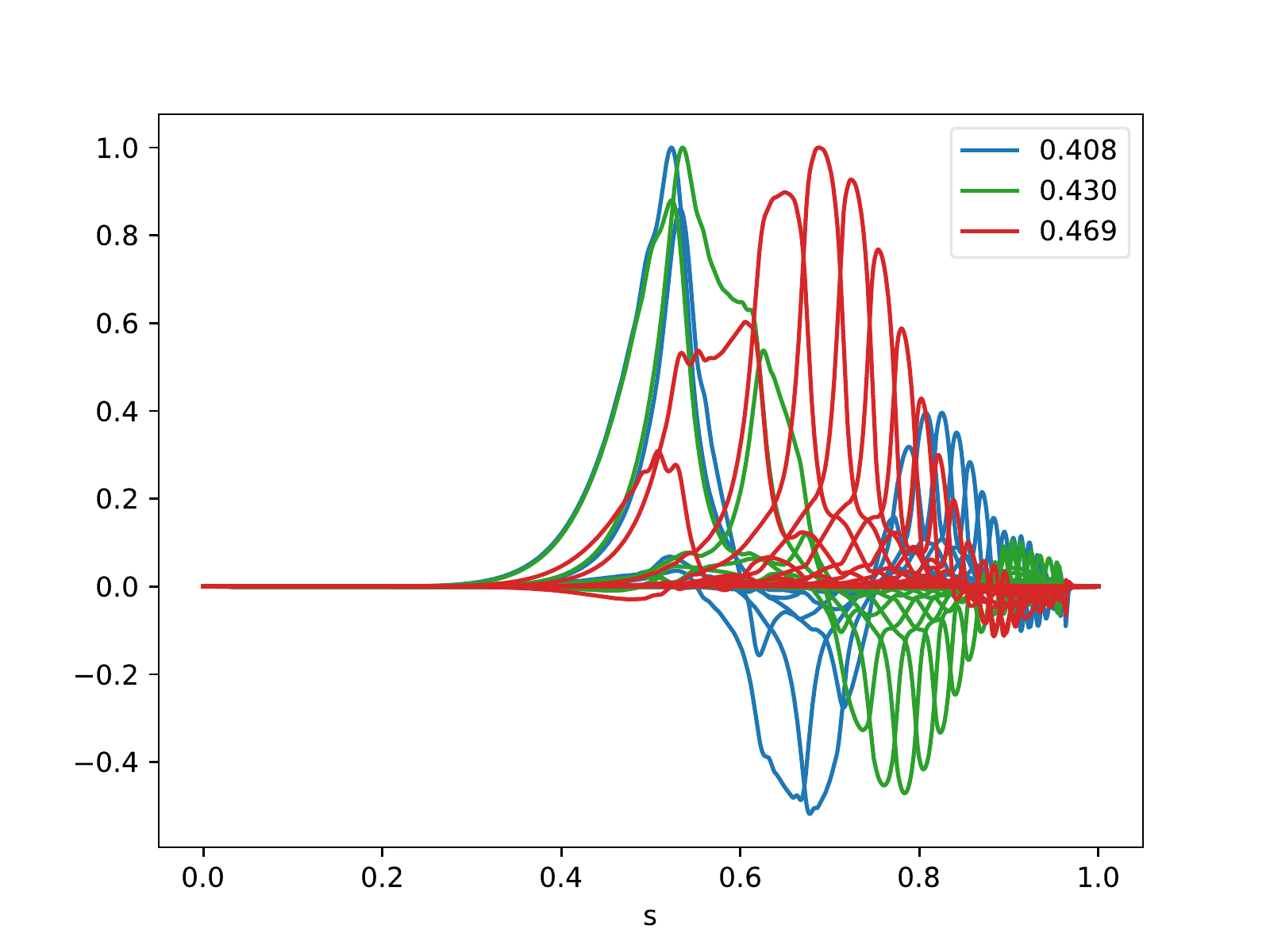}
	\caption{Radial harmonic structure for three TAE eigenfunctions found for $n=12$ with LIGKA, plotting the real part of the electrostatic potential. These eigenfunctions were calculated for use in reference~\cite{Schneller2016}. The legend refers to the frequencies of the modes, normalized to $\omegaA$.}
	\label{fig:n12_ligka_EFs}
\end{figure}

An example large-$n$ mode structure ($n=30$) is shown in figure~\ref{fig:n30_global_radial}.
In contrast, a low-$n$ mode ($n=12$) is shown in figure~\ref{fig:n12_global} (upper).
In the case of the low-$n$ mode, we observe a ``beating'' pattern in the evolution of the harmonics in figure~\ref{fig:n12_global} (lower), which is also observed in Fourier analysis.
This is further investigated, and LIGKA finds global 3 $n=12$ TAEs with slightly different continuum interaction and frequency variation $\approx15\%$.
These modes, shown in figure~\ref{fig:n12_ligka_EFs} are consistent with mode structures and beating patterns observed with ORB5.

If, however, we simulate this global $n=12$ TAE with a reduced annulus of $0.2 < s < 0.7$, which contains the dominant $q=12.5/12$ surface, then we instead no longer see mode beating, the signal contains only a single frequency, and the measured growth rate is different.
Given that the differences between the different modes occur due to the difference in the coupling across the outer domain, this therefore is not a surprising outcome, but it explains why a global treatment is found to be necessary to correctly describe global modes.

\begin{figure}
        \centering
        \includegraphics[width=0.36\textwidth]{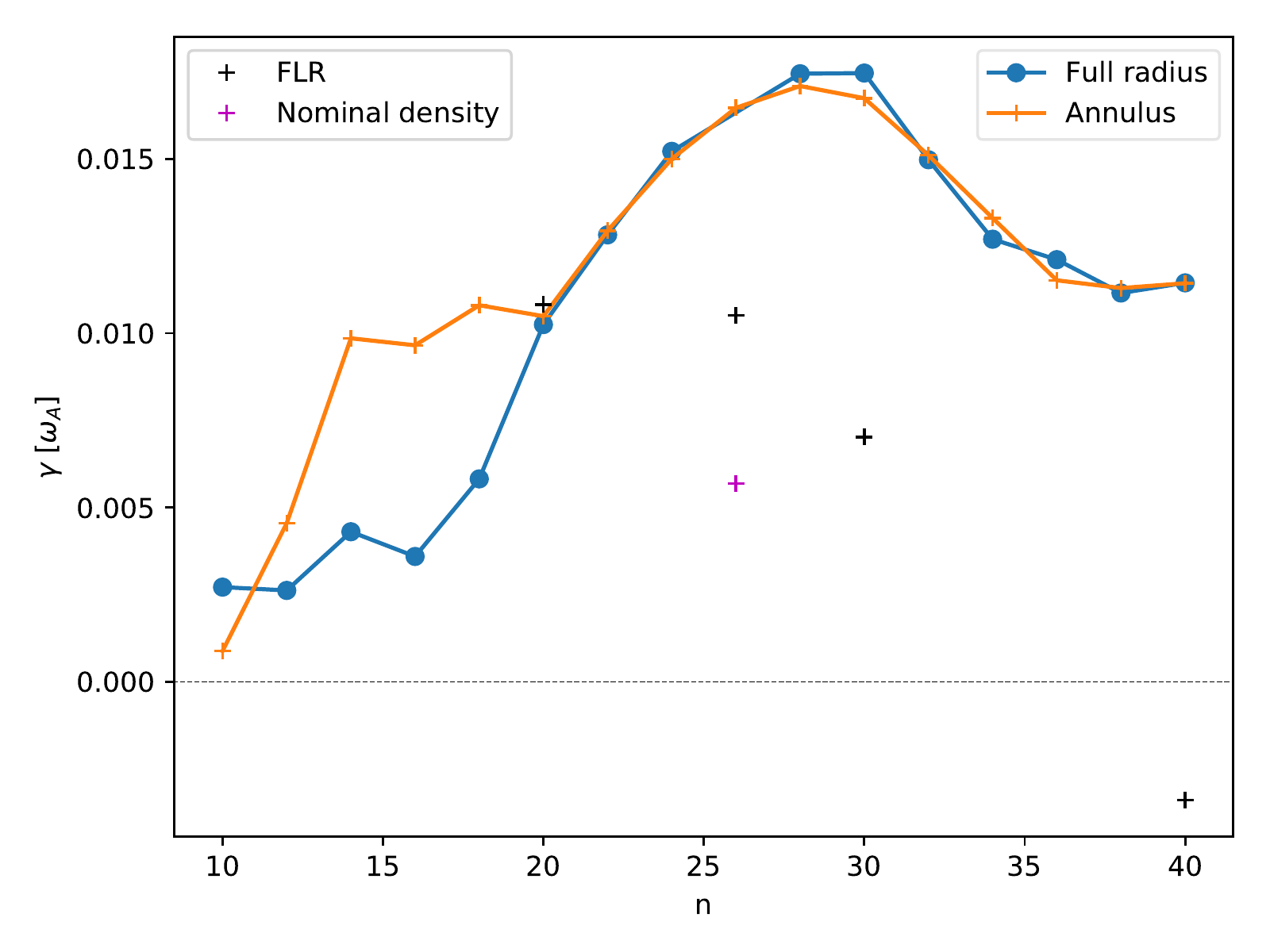}
        \includegraphics[width=0.36\textwidth]{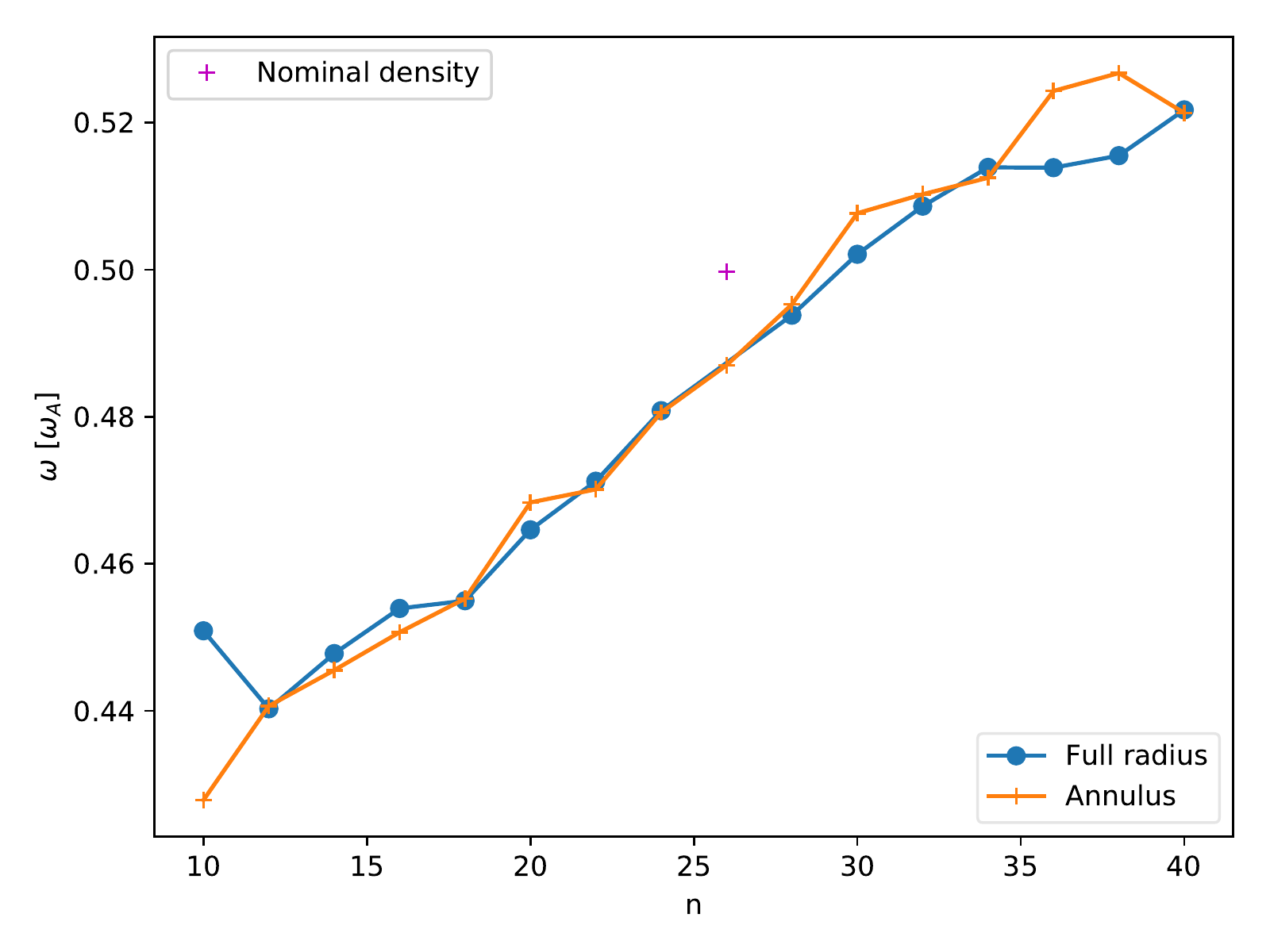}
        \caption{Scan of the mode growth rates (a) and frequencies (b) over a range of toroidal mode numbers $n$. These are calculated for simulations on both annular (crosses) and global (circles) domains. An additional point is added for reference, showing the results for $n=26$ with nominal EP density on the annular domain. In (a), we also add growth rates measured with finite Larmor radius (FLR) effects enabled in the EPs (black crosses), simulated on annular domains.}
        \label{fig:nscan_linear}
\end{figure}

Finally, we perform a scan of every 2nd $n$, ($10 \le n \le 40$) and plot the frequency and growth rates in figure~\ref{fig:nscan_linear} and we note the following observations.
Firstly, we observe a peak in the growth rate for n in the range $26$--$32$.
Secondly, for modes with $n\ge20$, the growth rate is independent of whether a full radial simulation was performed or one with $0.2 < s < 0.7$.
However, below $n=20$, we see that the annular simulations overestimate the growth rate for the reasons discussed above.

In addition to the points with double EP density and neglecting gyroaveraging, we include additional points at $n=\{20,26,30,40\}$ with FLR effects included, which can be seen to decrease the growth rate for $n>20$, push $n=40$ below marginal stability, and downshift the $n$ of the most unstable mode (from $\approx 28$ to $\approx 22$).
Finally, we include at $n=26$ also a point with nominal EP density.
We see in figure~\ref{fig:nscan_linear} (lower) that this point has an upshifted frequency with respect to the double EP density point.
This implies that non-pertubative EP effects are non-negligible for this case.

Finally, if one were to extrapolate towards a case with nominal EP density and FLR effects, then we would expect the most unstable modes to be weakly unstable, in rough agreement with the linear eigenvalue studies previously performed~\cite{Pinches_ITER_2015,Lauber_ITER_2015}.
However, any quantitative comparison should remember that the Maxwellian distribution function is expected to drive less strongly than a more realistic slowing down distribution for alpha particles, and the effect of the damping on the simplifications in the ion isotopes considered.

	\section{Nonlinear modelling}
\label{sec:nonlinear}
\begin{figure}
        \centering
        \includegraphics[width=0.4\textwidth]{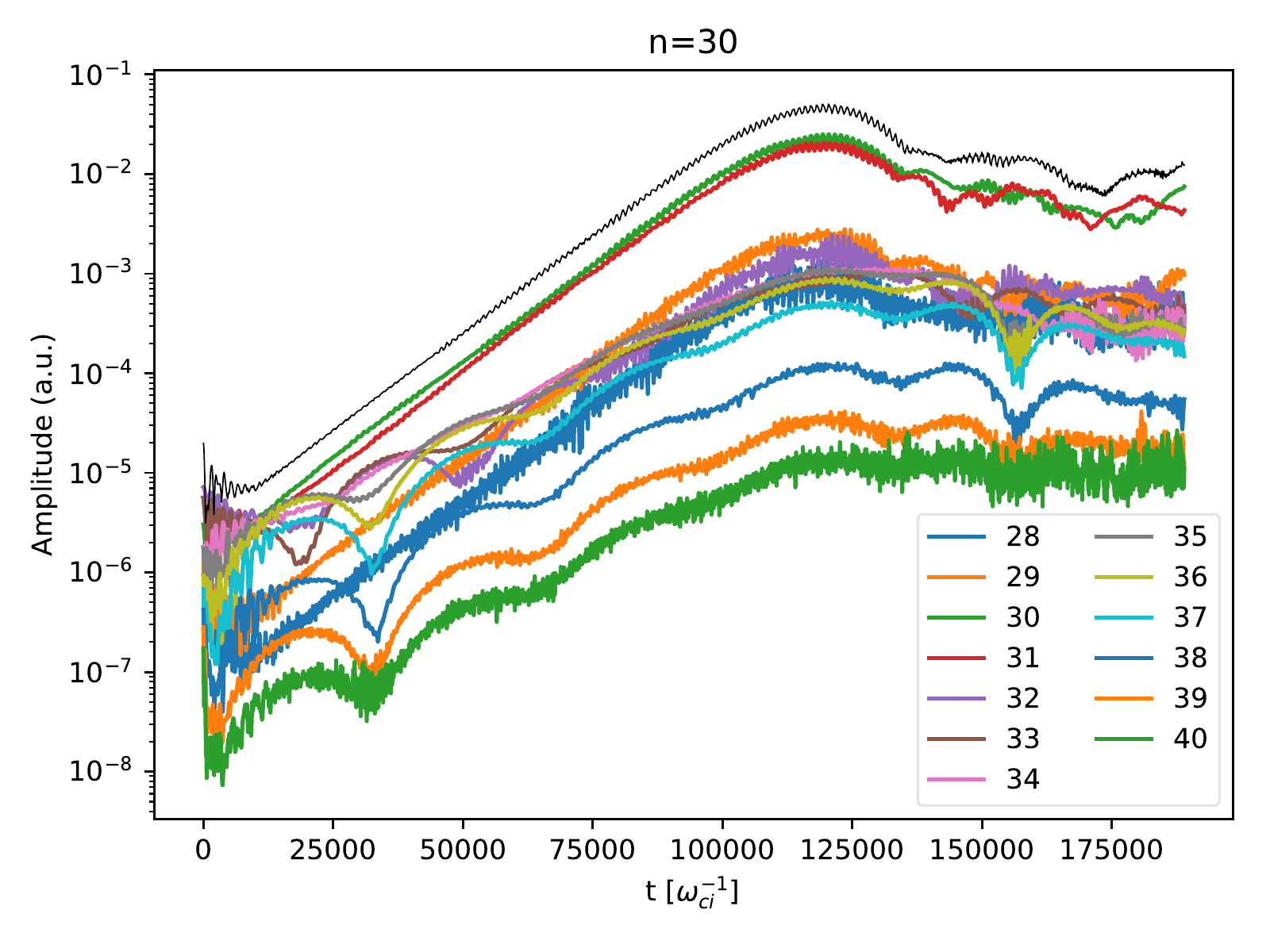}
        \caption{Nonlinear evolution of the poloidal harmonics of the electrostatic potential for $n=30$ TAEs. In black, the peak value of the toroidal envelope (LFS) is shown.}
        \label{fig:n30_nl_m_of_t_ann}
\end{figure}

\begin{figure}
        \centering
        \includegraphics[width=0.4\textwidth]{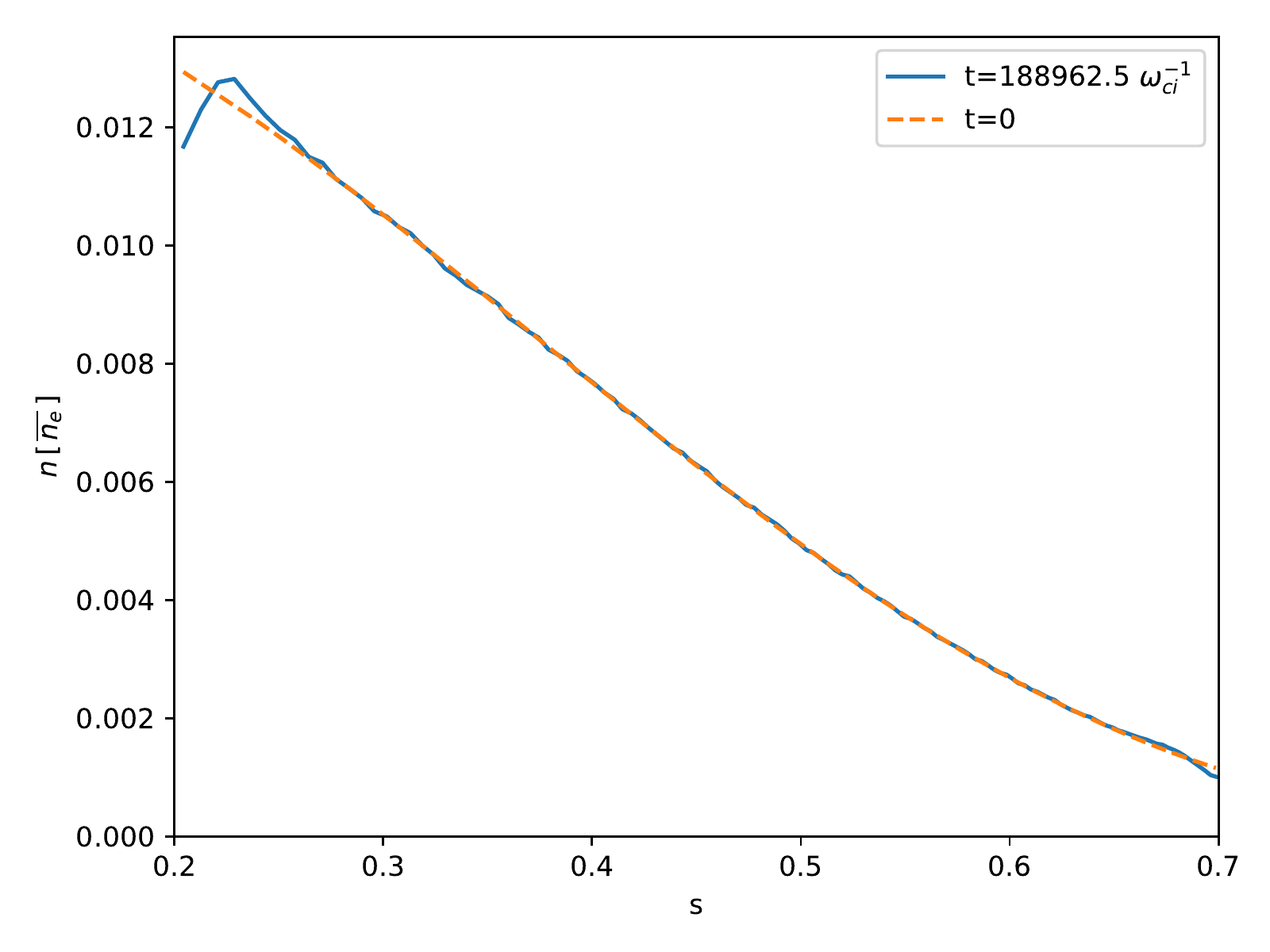}
        \caption{Initial and final profiles of the EP density for the simulation shown in figure~\ref{fig:n30_nl_m_of_t_ann}.}
        \label{fig:n30_nl_ann_den}
\end{figure}

Continuing from the linear results shown in the previous section, we now run longer simulations where the energetic particles are following non-perturbed trajectories\footnote{The results of \S\ref{sec:linear_stability} were also nonlinear simulations, but the analysis was done in the linear phase.}.
For the case of $n=30$, shown in figure~\ref{fig:n30_nl_m_of_t_ann}, we can see an example where the linear saturation of the TAE due to wave-particle trapping takes place.
The negligible redistribution of EPs by this mode is shown in figure~\ref{fig:n30_nl_ann_den}, where the maximum value of $\delta n_\trm{EP}(s)/n_\trm{EP}(s)$ is found to be $0.25\%$

\begin{figure}
        \centering
        \includegraphics[width=0.4\textwidth]{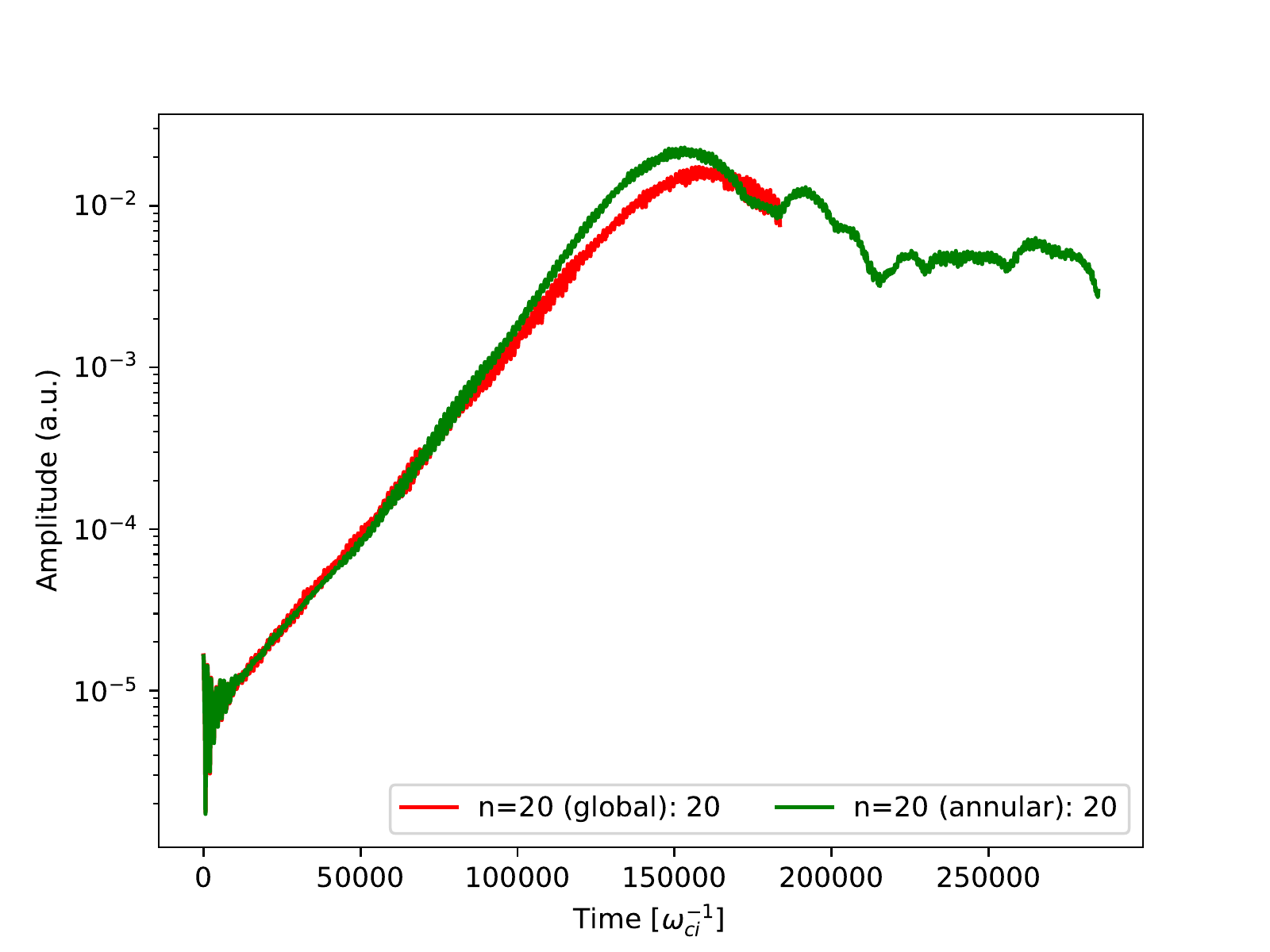}
        \caption{Nonlinear evolution of the toroidal envelopes (LFS) of the electrostatic potential for $n=20$ TAEs, simulated on full radius and annular domains.}
        \label{fig:n20_nl_n_of_t_comp}
\end{figure}

For $n=20$, the lowest mode number for which the linear simulations showed agreement between annular and full-radius simulations, we test this also nonlinearly, showing in figure~\ref{fig:n20_nl_n_of_t_comp} a comparison of the envelopes for annular and full radius simulations.
We observe a good agreement for $n=20$ (for $n=30$ there are no discernible differences).

\subsection{Multi-mode modelling}
\begin{figure}
        \centering
        \includegraphics[width=0.4\textwidth]{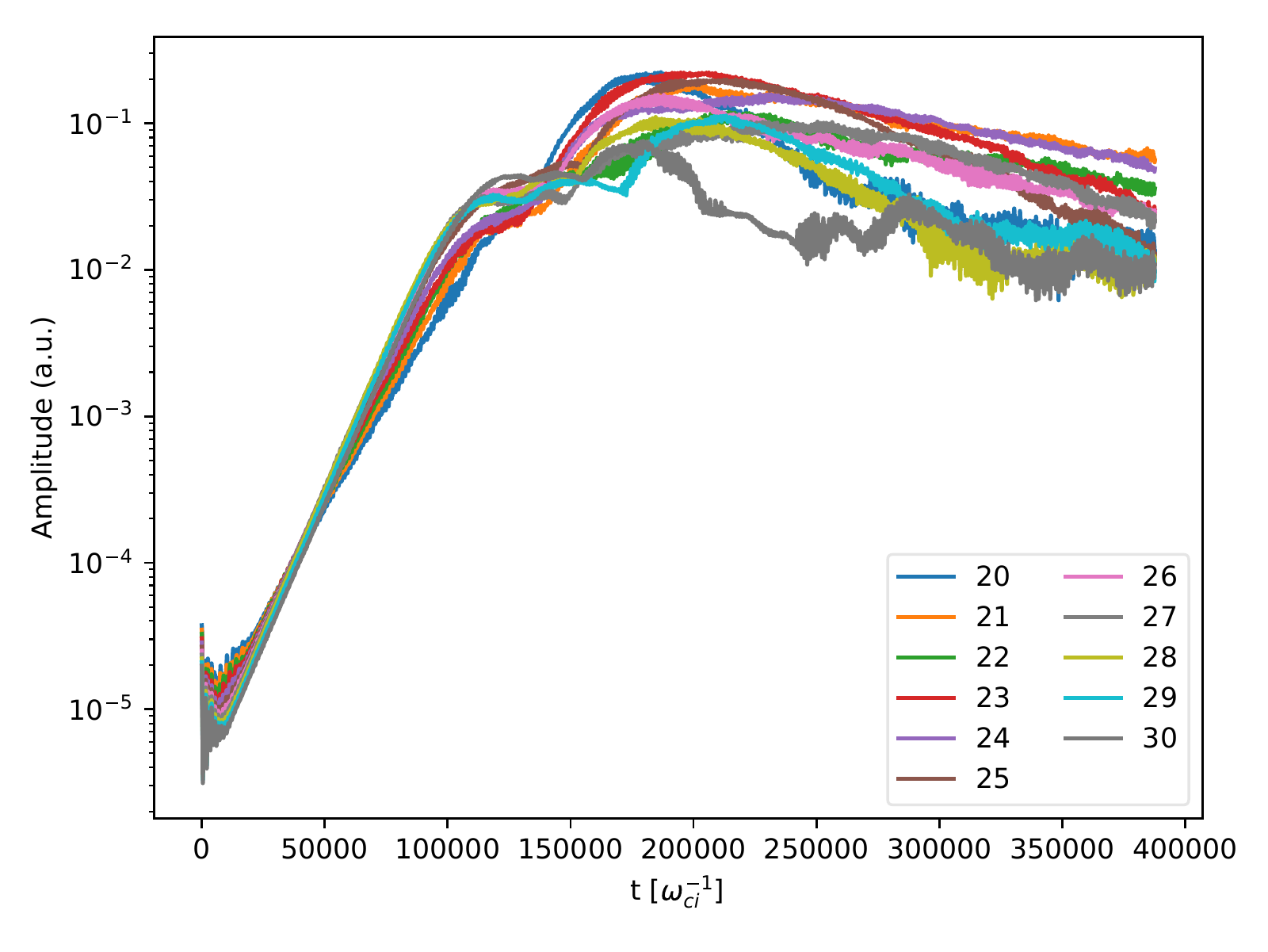}
        \caption{Nonlinear evolution of the peak values of the toroidal envelope of each toroidal mode $n$, simulated on an annular domain.}
        \label{fig:n20to30_nl_n_of_t_ann}
\end{figure}

\begin{figure}
        \centering
        \includegraphics[width=0.4\textwidth]{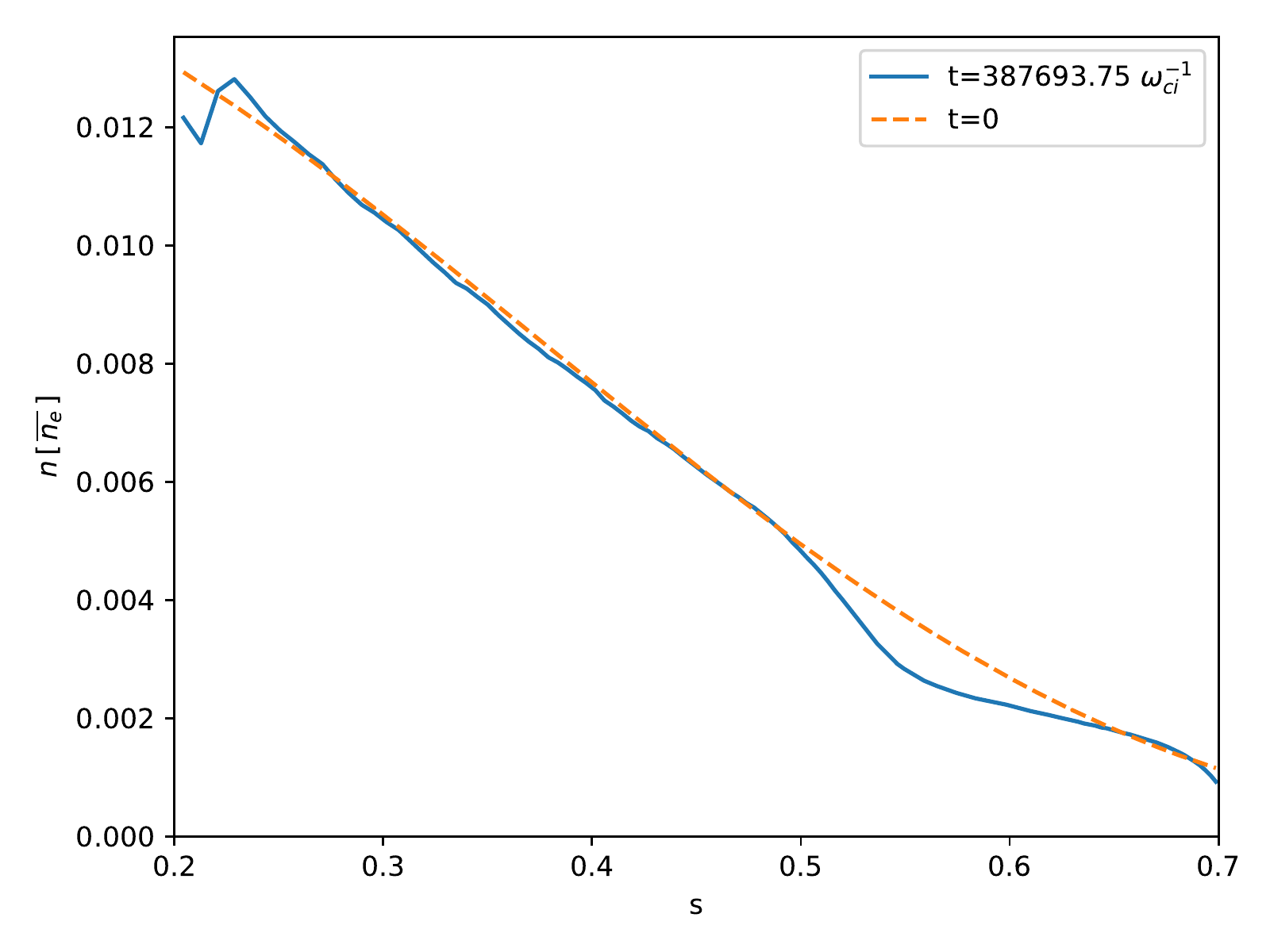}
        \caption{Initial and final profiles of the EP density for the simulation shown in figure~\ref{fig:n20to30_nl_n_of_t_ann}.}
        \label{fig:n20to30_nl_f_av_ann}
\end{figure}

Having demonstrated single-mode nonlinear simualtions with $n=20$ and $n=30$, we now perform a simulation with $n=\{20\ldots30$\}, using $10\times$ the number of markers.
The evolution of the mode envelopes is shown in figure~\ref{fig:n20to30_nl_n_of_t_ann}, where we see that the modes start to saturate in the range $100000 < t < 150000$ at similar amplitudes to the single-mode cases, but then are nonlinearly driven to ``another'' enhanced saturation at $t \approx 175000$, associated with the saturation of modes at outer radii. These modes are linearly subdominant and are only enhanced in the case of multiple modes by density redistribution.
This enhancement of the saturation amplitudes is an effect of the mode interaction, and leads to significantly more EP transport, as shown in figure~\ref{fig:n20to30_nl_f_av_ann}.
We see in this case that the modes overlap radially, giving a single large redistribution area~\cite{Berk1995a, Schneller2016}.
It was also found that with $n=\{20,22,\ldots,30\}$ that the final density profile essentially identical.
This implies that the double density case is quite far above the threshold over which the enhanced redistribution would be observed, as we observe resonance overlap even with the odd-n modes omitted.

It is important to note that the limitations outlined previously, most notably the enhanced EP density and the drift-kinetic EPs place a caveat on the results observed here, as both of these effects will decrease the mode drive.
However, the approximation of the alpha particles by a Maxwellian distribution is known to underestimate the growth rates.

	\section{Conclusions}
\label{sec:conclusions}
In this work, we have presented the first global initial value gyrokinetic simulations of TAEs in ITER.
In \S\ref{sec:numerical}, we began by demonstrating, through means of a study with the electron mass and the numerical time step, the relationship between the convergence of the measured linear growth rates with the time step and the the electron mass considered. We observed a weak dependence of the damping rate with respect to the electron mass, and a strong dependence of the numerical convergence properties on the electron mass. This study motivated and justified the choice of an electron mass $m_\trm{e}/m_\trm{i} = 1/200$ for the remainder of the work.
In \S\ref{sec:linear}, we demonstrated a variety of damped and driven \Alfv{} eigenmodes before showing the linear spectrum of TAEs, with their frequency and growth rates. We observe that the low-n global TAEs require a large domain to correctly resolve, or else we observe modes with incorrect growth rates.
In the nonlinear phase, we find that individual TAEs saturate at a level where the redistribution of EP density is small ($\delta n_\trm{EP}/n_\trm{EP} \approx 2.5\cdot 10^{-3}$). However, in the presence of multiple TAEs ($n=20\ldots30$), the redistribution becomes significant, due to the resonance overlap. We conclude that the system is well above the threshold for a transition to this enhanced saturation, as we observe the same redistribution in a case with only every second mode retained.

The obtained results are largely consistent with previous linear GK and nonlinear hybrid simulations~\cite{Pinches_ITER_2015, Schneller2016}: damping rates and nonlinear saturation behaviour for double nominal alpha particle density are found to agree well.
In addition, the role of EAEs and odd TAEs that have not been considered before has been clarified.
Based on this important step, more comprehensive EP studies with ORB5 and the related model hierarchy (LIGKA/HAGIS) can be carried out in order to address open issues such as more general distribution functions, scenario sensitivity studies and the interaction with the plasma turbulence.

	\section*{Acknowledgements} The authors would like to acknowledge discussions with A. Mishchenko, X. Wang, S. G\"unter, R. Hatzky, and A. Biancalani. The authors would also like to thank the members of the ORB5 team.
This work has been carried out within the framework of the EUROfusion Consortium and has received funding from the Euratom research and training program 2014-2018 and 2019-2020 under grant agreement No 633053. The views and opinions expressed herein do not necessarily reflect those of the European Commision.
Simulations presented in this work were performed on the MARCONI FUSION HPC system at CINECA and the HPC systems of the Max Planck Computing and Data Facility (MPCDF).

        \section*{References}
	\bibliographystyle{unsrt}
	\bibliography{twhs_nf20}
\end{document}